\documentclass[superscriptaddress,amsmath,amssymb, aps,prx,longbibliography,twocolumn, footinbib]{revtex4-2}

\usepackage{upgreek,mathtools}
\usepackage{color}
\usepackage{graphicx}
\usepackage{nccmath}
\usepackage{bm}
\usepackage{hyperref}
\hypersetup{colorlinks,breaklinks,
            urlcolor=[rgb]{0,0,0.36},
            linkcolor=[rgb]{0,0,0.36},
            citecolor=[rgb]{0,0,0.36}}
\usepackage[mathlines]{lineno}
\usepackage{dcolumn}
\usepackage{mathtools}  
\usepackage{dsfont}
\usepackage{comment}

\usepackage{epsfig,tabularx,multirow,booktabs}

\usepackage{braket}

\makeatletter
\renewcommand*\env@matrix[1][\arraystretch]{%
  \edef\arraystretch{#1}%
  \hskip -\arraycolsep
  \let\@ifnextchar\new@ifnextchar
  \array{*\c@MaxMatrixCols c}}
\makeatother

\definecolor{cadmiumgreen}{rgb}{0.0, 0.42, 0.24} 

\usepackage{pifont} 

\usepackage{adjustbox} 
\usepackage{tabularx} 
\newcolumntype{Y}{>{\centering\arraybackslash}X} 

\newcounter{SN}

\makeatletter
\renewcommand\footnotesize{%
   \@setfontsize\footnotesize\@ixpt{8}%
   \abovedisplayskip 8\p@ \@plus2\p@ \@minus4\p@
   \abovedisplayshortskip \z@ \@plus\p@
   \belowdisplayshortskip 4\p@ \@plus2\p@ \@minus2\p@
   \def\@listi{\leftmargin\leftmargini
               \topsep 4\p@ \@plus2\p@ \@minus2\p@
               \parsep 2\p@ \@plus\p@ \@minus\p@
               \itemsep \parsep}%
   \belowdisplayskip \abovedisplayskip
}
\makeatother

\usepackage{physics}

\newcommand{\Harvard}{Department of Physics, Harvard University, Cambridge, Massachusetts 02138, USA.}

\newcommand{\MaxP}{Max Planck Institute for the Structure and Dynamics of Matter, Luruper Chaussee 149, 22761 Hamburg, Germany.}

\newcommand{\BNL}{Condensed Matter Physics and Materials Science Division, Brookhaven National Laboratory, Upton, New York 11973, USA.}

\newcommand{\Oxford}{Clarendon Laboratory, University of Oxford, Parks Road, Oxford OX1 3PU, UK.}

\newcommand{\ETH}{Institute for Theoretical Physics, ETH Zurich, 8093 Zurich, Switzerland.}

\begin{document}

\title{ Principles of 2D terahertz spectroscopy of collective excitations:\\ the case of Josephson plasmons in layered superconductors}

\author{Alex~G\'{o}mez~Salvador}
\thanks{These authors contributed equally to this work.}
\affiliation{\ETH}
\author{Pavel~E.~Dolgirev}
\thanks{These authors contributed equally to this work.}
\affiliation{\Harvard}

\author{Marios~H.~Michael}
\thanks{These authors contributed equally to this work.}
\affiliation{\MaxP}

\author{Albert~Liu}
\thanks{These authors contributed equally to this work.}
\affiliation{\MaxP}
\affiliation{\BNL}

\author{Danica~Pavicevic}
\affiliation{\MaxP}

\author{Michael~Fechner}
\affiliation{\MaxP}

\author{Andrea~Cavalleri}
\affiliation{\MaxP}
\affiliation{\Oxford}

\author{Eugene~Demler}
\affiliation{\ETH}

\begin{abstract}
    Two-dimensional terahertz spectroscopy (2DTS), a terahertz analogue of nuclear magnetic resonance, is a new technique poised to address many open questions in complex condensed matter systems. The conventional theoretical framework used ubiquitously for interpreting multidimensional spectra of discrete quantum level systems is, however, insufficient for the continua of collective excitations in strongly correlated materials. Here, we develop a theory for 2DTS of a model collective excitation, the Josephson plasma resonance in layered superconductors. Starting from a mean-field approach at temperatures well below the superconducting phase transition, we obtain expressions for the multidimensional nonlinear responses that are amenable to intuition derived from the conventional single-mode scenario. We then consider temperatures near the superconducting critical temperature $T_c$, where dynamics beyond mean-field become important and conventional intuition fails.  As fluctuations proliferate near $T_c$, the dominant contribution to nonlinear response comes from an optical parametric drive of counter-propagating Josephson plasmons, which
   gives rise to 2D spectra that are qualitatively different from the mean-field predictions. As such, and in contrast to one-dimensional spectroscopy techniques, such as third harmonic generation, 2DTS can be used to directly probe thermally excited finite-momentum plasmons and their interactions. Our theory provides a clear interpretation of recent 2DTS measurements on cuprates, and we discuss implications beyond the present context of Josephson plasmons.
\end{abstract}

\date{\today}

\maketitle

\section{Introduction}

Since its inception over half a century ago, nuclear magnetic resonance (NMR)~\cite{ernst1990principles,friebolin1991basic} has not only become a standard tool in the fundamental sciences for resolving structure and interactions in both molecular~\cite{book_Zerbe} and solid-state systems~\cite{Reif2021}, but has also become indispensable to modern technologies as diverse as magnetic resonance imaging~\cite{book_McRobbie} and plant analysis \cite{Kim2010}. In recent years, optical analogues of nuclear magnetic resonance that interrogate the electronic constituents of matter, termed multidimensional coherent spectroscopies~\cite{hamm_zanni_2011,Cundiff2013}, have come to the fore. With unique capabilities to, for example, identify coupling between different resonances, disentangle homogeneous and inhomogeneous broadening mechanisms, and resolve energy transfer pathways, these techniques have revolutionized our understanding of complex atomic~\cite{Bruder2019}, chemical \cite{Mukamel2009}, and biological \cite{Petti2018} systems, with growing applications for condensed matter systems \cite{Cundiff2012}. However, the potential of these techniques extends far beyond these cases for which they were originally envisioned. 

Indeed, the characteristics of systems in which these multidimensional spectroscopies excel, disordered systems with numerous interacting degrees of freedom, are shared by strongly correlated `quantum materials’ \cite{Keimer2017}. Loosely defined, quantum materials comprise systems for which even a qualitative understanding of their properties requires a quantum mechanical treatment. Discerning their complex nature and harvesting their unique properties holds tremendous promise for future technologies \cite{Giustino2021}. Many open questions remain in regards to these materials, both to the physical phenomena underlying their properties and to practical applications, including rational design and device implementation. Some of these questions, while impervious to conventional probes, could be addressed by multidimensional techniques. Yet the typical low energy scales of collective excitations in such systems have hampered previous efforts. These energy scales are often in the terahertz optical domain \cite{Nicoletti2016}, and the traditional challenge of generating strong, coherent radiation at these frequencies is well-known as the `terahertz gap' \cite{book_Pavlidis}. Fortunately, this terahertz gap is now closing and recent developments in intense low-frequency light sources \cite{Nicoletti2016,Fulop2020} have enabled so-called two-dimensional terahertz spectroscopies (2DTS) \cite{Lu2019,Reimann2021} that probe quantum materials on their fundamental energy scales.

In recent years, an increasing number of groups have applied 2DTS to a wide range of condensed matter systems, ranging from superconductors \cite{Mootz2022,Luo2023,Liu_2023_echo} to ferroics \cite{Lu2017,Lin2022,zhang2023terahertz}, and even topological materials \cite{Blank2023}. However, the interpretation of their 2DTS spectra often relies on intuition derived from localized, discrete, single-particle excitations in atoms, molecules, and other quantum-confined systems \cite{Kuehn2009}. This is fundamentally different from delocalized, collective excitations, characterized by a continuous dispersion. Upon recognizing this point, many questions arise concerning their optical nonlinearities. How do we transition from the nonlinear optical response of discrete quantum states to an energy band continuum? What are the signatures of parametric creation and annihilation processes \cite{Michael2022,dolgirev2022periodic}, and the correlations that result? What is the effect of an interface that spectrally shapes the optical response? These and other salient questions have yet to be addressed in a unified manner, which calls for a reformulation of the ubiquitious theoretical framework used to interpret conventional multidimensional spectra \cite{hamm_zanni_2011,book_MDCS,Mukamel2009}. Here, we address these questions theoretically using the well-known Josephson plasma resonance in layered superconductors \cite{Laplace2016} as a model collective excitation that exhibits strong sine-Gordon nonlinearities. This choice is motivated by recent experimental reports applying 2DTS to the layered cuprate superconductor La$_{2-x}$Sr$_x$CuO$_4$~\cite{Liu_2023_echo,albert_phase_transition}, as well as the fundamental and technological importance of high-$T_c$ cuprates. While Josephson plasmons represent a specific system, their electrodynamics is quite generic~\cite{Savelev2010}, allowing us to infer a broader insight into 2D spectroscopy of collective excitations (expanded upon in Appendix~\ref{appendix:nonlinear_Gen}). 

This paper presents three main results. The first is a derivation of the third order optical response of Josephson plasmons within mean-field theory (see Sec.~\ref{section_2} and Eq.~\eqref{eqn:chi_3_gen}), applicable  at low temperatures $T\ll T_c$ deep in the superconducting phase. This response exhibits both analogies and important differences to the optical properties of quantum level systems -- one such difference is the presence of air-superconductor interface which we carefully take into account via the Fresnel formalism. We find that the Josephson plasmon optical nonlinearity Eq.~\eqref{eqn:chi_3_gen} descends from the linear optical response, a feature that we argue is generic to mean-field approaches (see also Appendix~\ref{appendix:nonlinear_Gen}).

Our second result is detailed in Sec.~\ref{section_3}, where we introduce the technique of 2DTS and present expressions for the measured multidimensional nonlinear responses (Eqs.~\eqref{eqn:map_AAB} and~\eqref{eqn:map_ABB}). We discuss additional important considerations that, while not relevant for quantum level systems, need to be taken into account when analysing and designing 2DTS measurements of complex solids. Simulations of 2D THz spectra are then presented and resultant two-dimensional lineshapes are carefully analyzed. These lineshapes are found to remarkably preserve information about line-broadening mechanisms naively expected from a single-mode picture. We remark that revealing these mechanisms is among the most useful and unique capabilities of 2DTS. 

Finally in Sec.~\ref{sec:squeeze}, we go beyond mean-field theory and consider the scenario of optical nonlinearities near a phase transition $T\lesssim T_c$. In this regime fluctuations proliferate and therefore can no longer be neglected -- their correction within the Gaussian approximation to the nonlinear susceptibility is encoded in Eq.~\eqref{eqn:chi_3_sq}. The key nonlinear process that one needs to take into account as $T$ increases is an optical parametric drive of counter-propagating Josephson plasmons~\cite{dolgirev2022periodic,Gabriele2021} -- this process results in characteristic 2D maps qualitatively different from the mean-field ones, thereby allowing us to isolate the non-mean-field nonlinearities.
The resulting 2DTS signatures, absent in the linear optical response, defy intuition derived from conventional spectroscopy of dispersionless quantum level systems. We argue that 2DTS, in contrast to one-dimensional nonlinear spectroscopy techniques such as third harmonic generation, allows us to probe finite-momentum thermal fluctuations and their interactions.
Our theory is validated by, and provides a natural interpretation of recent 2DTS measurements reported in Ref.~\cite{albert_phase_transition}. Implications beyond the context of Josephson plasmonics are then discussed.

\section{Nonlinear Mean-Field Response of layered superconductors } \label{section_2}

One of the key aspects of layered cuprate superconductors is the strong anisotropy which renders the $c$-axis Josephson plasma modes to be primary low-energy collective excitations in the system \cite{Gabriele2022}, with frequencies typically in the terahertz range. Motivated by recent experimental developments in the field of 2DTS~\cite{Liu_2023_echo,albert_phase_transition}, which allows one to study nonlinearities of the Josephson plasmons in cuprates, here we develop the theory of nonlinear electrodynamics of layered superconductors, which extends the framework of Ref.~\cite{PhysRevB.50.12831} to account for the air-superconductor interface (see Fig.~\ref{fig:1}).

Specifically, analysis in this section is based on mean-field theory, valid for $T\ll T_c$. The central result here is Eq.~\eqref{eqn:chi_3_gen}, which relates the third-order susceptibility $\chi^{(3)}(\omega_1,\omega_2,\omega_2)$ to the linear optical response of the system.  While the relation~\eqref{eqn:chi_3_gen} was derived for a specific model, it is argued to be generic to nonlinear many-body systems that can be describable via mean-field equations of motion (see Appendix~\ref{appendix:nonlinear_Gen}). 

In the following sections we will analyse 2DTS in detail, as it directly probes $\chi^{(3)}$. We will also go beyond the mean-field analysis and consider the plasma squeezing mode, expected to play a prominent role near the transition temperature. Throughout the paper, we discuss the additional information 2DTS measurements bring compared to, for instance, linear optical spectroscopy and nonlinear third harmonic generation.

\subsection{Equations of motion}

To describe the nonlinear electromagnetic response of layered superconductors we employ the two-fluid model developed in Ref.~\cite{PhysRevB.50.12831}. The layered superconductivity is described using the Lawrence-Doniach model~\cite{Bul_vortex}, which neglects the dynamics of the order parameter amplitude $\Delta$ because it assumes that $\omega\ll \Delta$, where $\omega$ is the typical probe frequency. In other words, the superconducting response is fully encoded in the dynamics of the order parameter phase $\varphi_n(\bm r,t)$, where $n$ is the layer index and $\bm r$ is the in-plane coordinate. The normal fluid is described via phenomenological Ohm's law. As such, 
within this two-fluid model, the $z$-axis current density between layers $n$ and $n+1$ is given by:
\begin{align}
    J_{z;n,n+1} = J_0 \sin \varphi_{n,n+1} + \sigma_0 E_{z; n,n+1},\label{eqn:J_z_v0}
\end{align}
where $J_0$ is the Josephson critical current and 
\begin{align}
    \varphi_{n,n+1} = \varphi_n - \varphi_{n + 1}  - \frac{2\pi}{\Phi_0} \int_{n s}^{(n + 1) s} dz\, A_z
\end{align}
is the gauge-invariant phase difference between layers $n$ and $n + 1$. Here $s$ is the distance between adjacent layers, $\sigma_0$ is the $z$-axis normal conductivity (for an additional discussion about $\sigma_0$, see Ref.~\cite{PhysRevB.50.12831}), and $\Phi_0 = 2\pi c/(2 e)$ (throughout the paper, we set $\hbar = k_B = 1$). We also defined:
\begin{align}
    E_{z; n,n+1} = \frac{1}{s}\int_{n s}^{(n + 1) s} dz\, E_z,
\end{align}
etc. The two constituent fluids are coupled to each other via Maxwell's equations, which ensure that the Coulomb screening effects are taken into account. When working with the electromagnetic field, we employ the gauge for which the scalar potential is zero, i.e., $\bm E(\bm r,z,t) = - \partial_t \bm A(\bm r,z,t)/c$ and $\bm B(\bm r,z,t) = \nabla \times \bm A(\bm r,z,t)$, with $c$ being the speed of light and $\bm A$ being the vector potential. Inside the sample, we have:
\begin{gather}
    \varepsilon_{\infty} \nabla \cdot \bm E(\bm r,z,t) = 4 \pi \rho(\bm r,z,t),\\
    \nabla \times \bm B(\bm r,z,t) = \frac{\varepsilon_{\infty}}{c}\partial_t \bm E(\bm r,z,t) + \frac{4\pi}{c} \bm J(\bm r,z,t),
\end{gather}
where $\rho$ and $\bm J$ are the three-dimensional charge and current densities, respectively, related to each other via the continuity relation; $\varepsilon_{\infty}$ is the high-frequency $c$-axis dielectric constant.
For the in-plane current densities, we write the London relation
\begin{align}
    J_n(\bm r) = - J_0\varkappa^2 s \Big[ \nabla \varphi_n(\bm r) + \frac{2\pi}{\Phi_0}\bm A_n(\bm r) \Big],
\end{align}
where $\varkappa \gg 1$ is the anisotropy parameter. We also write the Josephson relation
\begin{align}
    \partial_t \varphi_{n,n+1} = 2 e sE_{z;n, n+ 1}, \label{eqn:JR}
\end{align}
expected to hold at low temperatures when one can neglect the presence of pancake vortices. 

Combining all of the above equations and in the limit $(\omega \lambda_{ab}/c)^2\ll 1$, where $\lambda_{ab}$ is the London penetration depth for the in-plane currents, one obtains~\cite{PhysRevB.50.12831,koshelev1999fluctuation} ($\psi_n \equiv \varphi_{n,n+1}$):
\begin{align}
    ( \partial_t^2 + \gamma \partial_t ) \psi_n - \nabla^2 L_{nm}\psi_m + \Lambda_0 \sin\psi_n = 0,
    \label{eqn: Bulaevskii}
\end{align}
with $\gamma = 4\pi \sigma_0/\varepsilon_\infty$, $\Lambda_0 = c_0^2 /\lambda_J^2$, and
\begin{align}
    L_{nm} = \frac{c_0^2}{N} \sum_{k} 
    \frac{e^{ik(n - m)}}{ 2(1 - \cos k) + s^2/\lambda_{ab}^2}.
\end{align}
Here $N$ is the total number of layers ($ k = 2\pi n/N$, with $n$ being integer), $\lambda_J= \varkappa s$, and  $c_0 = c s/(\lambda_{ab}\sqrt{\epsilon_{\infty}}) $ is the Swihart velocity.

\subsection{Boundary conditions}

In optical experiments, both linear and nonlinear, one sends light onto the superconducting sample and then measures, for instance, the reflected light (see Fig.~\ref{fig:1}). In practice, to evaluate the latter, one can separately solve Maxwell's equations in the air and the material, Eq.~\eqref{eqn: Bulaevskii}, and then match the solutions using the Fresnel boundary conditions. We assume i) normal incidence and ii) the incident light is homogeneous along the $yz$-plane with the electric field being parallel to the $z$-axis (see Fig.~\ref{fig:1}). As such, the boundary conditions at $x = 0$ are given by:
\begin{align}
      E^{\rm in}_z(y,t) + E^{\rm r}_z(y,t) & = E^{\rm t}_z(y,t), \label{eqn:el_bc}
      \\     
    B_y^{\rm in}(y,t) + B_y^{\rm r}(y,t) & = B_y^{\rm t}(y,t).\label{eqn:mag_bc}
\end{align}
Here, the superscripts r and t refer to the incoming, reflected, and transmitted light, respectively. For future reference, we note that, due to the homogeneity of the incoming light pulses along the $z$-axis, the gauge-invariant phase difference $\psi_n = \psi$ does not depend on the layer index $n$.

Inside the sample, the Josephson relation~\eqref{eqn:JR} is understood as $E^{\rm t}_z  = \partial_t \psi/(2es)$. For the magnetic field, one generically has~\cite{PhysRevB.50.12831}
\begin{align}
    \hat{z}\times \bm B_{n,n+1} = \frac{4\pi \lambda^2_{ab}}{c s} (\bm J_{n + 1} - \bm J_n) - \frac{\Phi_0}{2\pi s}\nabla \varphi_{n,n+1}.
\end{align}
However, the $z$-axis homogeneity of the incoming radiation implies that $\bm J_{n + 1} = \bm J_n$ so that
\begin{gather}
   \bm B^{\rm t} = \frac{\Phi_0}{2\pi s} \hat{z} \times \nabla \psi.
   \label{eqn:mag_phase}
\end{gather}

For the normal incidence we consider here, the boundary conditions~\eqref{eqn:el_bc} and~\eqref{eqn:mag_bc} further simplify to:
\begin{align}
    \frac{1}{4es}\partial_t\psi - \frac{\Phi_0}{4\pi s}\partial_x \psi = E^{\rm in}_z(x=0,t). \label{eqn:bc_main}
\end{align}
This result follows from relations~\eqref{eqn:JR} and~\eqref{eqn:mag_phase}, and the fact that in the air we have
\begin{align}
     B_y^{\rm in}(\bm r, t) = - E_z^{\rm in}(\bm r, t),\quad B_y^{\rm r}(\bm r, t) =  E_z^{\rm r}(\bm r, t).
     \label{eqn:EB_relations_air}
\end{align}
Equation~\eqref{eqn:bc_main} is particularly useful as it directly relates the incoming light to how it affects the dynamics inside the sample.
The reflected light, which encodes the primary observable of interest below, is then given by:
\begin{align}
    E^{\rm r}(x=0,t) = \frac{1}{4es}\partial_t\psi + \frac{\Phi_0}{4\pi s}\partial_x \psi.
\end{align}

\subsection{Nonlinear third-order susceptibility}

We now assume that the incident light is weak enough so that we may carry out a perturbative analysis in $E^{\rm in}$, but strong enough so that the resulting nonlinearities are measurable. Hence, we write: 
\begin{align}
    \psi(x,t) = \psi^{(1)}(x,t) + \psi^{(3)}(x,t) + \dots
\end{align}
We limit ourselves to the third-order response $\psi^{(3)}(x,t)$, as the second-order response is zero. From Eq.~\eqref{eqn: Bulaevskii}, we obtain the following coupled set of equations:
\begin{gather}
    (\partial_t^2 + \gamma\partial_t - L \partial_x^2 + \Lambda_0)\psi^{(1)} = 0,\label{eqn:1d_SG_psi1}\\
    (\partial_t^2 + \gamma\partial_t - L \partial_x^2 + \Lambda_0)\psi^{(3)} = \frac{\Lambda_0}{6}\left(\psi^{(1)}\right)^3, \label{eqn:1d_SG_psi3}
\end{gather}
where $L = \sum_m L_{nm} = c^2/\varepsilon_\infty$. As follows from Eq.~\eqref{eqn:bc_main}, the boundary conditions for $\psi^{(1)}$ and $\psi^{(3)}$ now read
\begin{gather}
    \partial_t \psi^{(1)}(0,t) - c\partial_x\psi^{(1)}(0,t) = 2{\cal E}^{\rm in}(t),\label{eqn:bc_1d_psi1}\\
    \partial_t \psi^{(3)}(0,t) - c\partial_x\psi^{(3)}(0,t) = 0,\label{eqn:bc_1d_psi3}
\end{gather}
where we have defined for notational convenience $\mathcal{E}^{\rm in}(t) = 2es E^{\rm in}(x=0,t)$. The leading term $\psi^{(1)}(x,t)$ describes the linear response which then acts as a drive for $\psi^{(3)}(x,t)$.

Since $\psi^{(1)}(x,t)$ satisfies the linear wave equation~\eqref{eqn:1d_SG_psi1}, one can generically write:
\begin{align}
    \psi^{(1)}(x,t) = \int \frac{d\omega}{2\pi}\Tilde{\psi}^{(1)}(\omega) e^{ik_x(\omega)x -i\omega t},
\end{align}
where 
\begin{align}
    L k_x^2(\omega) = \omega^2 + i\gamma \omega - \Lambda_0. \label{eqn:1d_kx}
\end{align}
The root  $k_x(\omega)$ in Eq.~\eqref{eqn:1d_kx} is chosen such that it corresponds to waves propagating away from the surface with $\text{Im}\, k_x(\omega) > 0$. The amplitudes $\Tilde{\psi}^{(1)}(\omega)$ are found using the boundary condition~\eqref{eqn:bc_1d_psi1} at $x = 0$:
\begin{align}
    \Tilde{\psi}^{(1)}(\omega) = \frac{2 i {\cal E}^{\rm in}(\omega)}{\omega + c k_x(\omega)} = \frac{i}{\omega}t(\omega)\mathcal{E}^{\rm in}(\omega). \label{eqn:lin_resp}
\end{align}
Here $t(\omega) = 2/(1+\sqrt{\epsilon(\omega)})$ is nothing but the transmission coefficient with $\sqrt{\epsilon(\omega)} = ck_x(\omega)/\omega$. Indeed, using the Josephson relation~\eqref{eqn:JR}, one can recover the usual Fresnel transmission relation given by: 
\begin{equation}
    E^{\rm t}(x=0,\omega) = \frac{2}{1+\sqrt{\epsilon(\omega)}}E^{\rm in}(x=0,\omega).
\end{equation}

Having determined the leading harmonic $\psi^{(1)}(x,t)$, we turn to compute $\psi^{(3)}(x,t)$. Since Eq.~\eqref{eqn:1d_SG_psi3} is linear as well, one can generically write $\psi^{(3)}(x,t)$ as:
\begin{align}
    \psi^{(3)}(x,t) = \int \frac{d\omega}{2\pi}\Tilde{\psi}^{(3)}(\omega) e^{ik_x(\omega)x -i\omega t} + \psi_{3}^{\rm dr}(x,t),
\end{align}
where the first term encodes the generic solution of the homogeneous part of Eq.~\eqref{eqn:1d_SG_psi3} and $\psi^{(3)}_{\rm dr}(x,t)$ is given by:
\begin{widetext}
\begin{align}
    \psi^{(3)}_{\rm dr}(x,t) & = \frac{\Lambda_0}{6}\int \frac{d\omega}{2\pi}\int \frac{d k}{2\pi} \frac{e^{ikx - i\omega t}}{-\omega^2 - i\gamma \omega + L k^2 + \Lambda_0  } \int \frac{d\omega_1}{2\pi}\frac{d\omega_2}{2\pi}\frac{d\omega_3}{2\pi} 2\pi \delta(\omega_1 + \omega_2 + \omega_3 - \omega) \notag\\
    &
    \times 2\pi \delta(k_x(\omega_1) + k_x(\omega_2) + k_x(\omega_3) - k) \Tilde{\psi}^{(1)}(\omega_1)\Tilde{\psi}^{(1)}(\omega_2)\Tilde{\psi}^{(1)}(\omega_3).
\end{align}
The amplitudes $\Tilde{\psi}^{(3)}(\omega)$ are obtained from the boundary condition~\eqref{eqn:bc_1d_psi3}:
\begin{align}
    \Tilde{\psi}^{(3)}(\omega)  & = \frac{-\omega \psi^{(3)}_{\rm dr}(x = 0,\omega) + ic \partial_x\psi^{(3)}_{\rm dr}(x = 0,\omega) }{\omega + ck_x(\omega)} = -
    \frac{\Lambda_0}{6}\int \frac{d k}{2\pi} \frac{1}{-\omega^2 - i\gamma \omega + L k^2 + \Lambda_0  } \frac{ \omega + c k}{\omega + c k_x(\omega)} \notag\\
    &
    \times \int \frac{d\omega_1}{2\pi}\frac{d\omega_2}{2\pi}\frac{d\omega_3}{2\pi} 2\pi \delta(\omega_1 + \omega_2 + \omega_3 - \omega)  2\pi \delta(k_x(\omega_1) + k_x(\omega_2) + k_x(\omega_3) - k) \Tilde{\psi}^{(1)}(\omega_1)\Tilde{\psi}^{(1)}(\omega_2)\Tilde{\psi}^{(1)}(\omega_3).
\end{align} 
For ease of notations, we define $\bar{\omega} = \omega_1+\omega_2+\omega_3$ and $k_x(\omega_1,\omega_2,\omega_3) = k_x(\omega_1) + k_x(\omega_2) + k_x(\omega_3)$.
For the reflected light at $x =0$, we then get:
\begin{align}
    \mathcal{E}_{\rm r}^{(3)}(t)&  = \frac{\partial_t \psi^{(3)}(0,t) + c\partial_x \psi^{(3)}(0,t)}{2}
    = \int\frac{d\omega_1}{2\pi}\int\frac{d\omega_2}{2\pi}\int\frac{d\omega_3}{2\pi}  e^{-i\bar{\omega} t} \chi^{(3)}(\omega_1,\omega_2,\omega_3) \mathcal{E}^{\rm in}(\omega_1)\mathcal{E}^{\rm in}(\omega_2)\mathcal{E}^{\rm in}(\omega_3).
\end{align}
\end{widetext}
Notably, we find that the third-order susceptibility
\begin{align}
    \chi^{(3)}(\omega_1,\omega_2,\omega_3) &= \frac{\Lambda_0\epsilon_\infty}{12}\frac{1}{c k_x(\omega_1,\omega_2,\omega_3)/\bar{\omega} + \sqrt{\epsilon(\bar{\omega})}  }\notag \\ &\times \frac{t(\bar{\omega})}{\bar{\omega}} \frac{t(\omega_1)}{\omega_1} \frac{t(\omega_2)}{\omega_2} \frac{t(\omega_3)}{\omega_3} \label{eqn:chi_3_gen}
\end{align}
is expressed solely through the dielectric function $\epsilon(\omega)$, which is fully determined by the linear response function~\eqref{eqn:lin_resp}. Such factorization holds only within mean-field approximation; nevertheless, the form in Eq.~\eqref{eqn:chi_3_gen} seems to be generic to any classical nonlinear reflectivity problem (see also Appendix~\ref{appendix:nonlinear_Gen}). We interpret various terms entering in Eq.~\eqref{eqn:chi_3_gen} as follows: i) the last three factors of $t(\omega_i)$ take into account the three transmission coefficients for the three incoming electric fields, ii) the term in the first row of Eq.~\eqref{eqn:chi_3_gen} encodes the dynamics of the 3-wave-mixed phase, and iii) the factor of $t(\bar{\omega})$ accounts for the transmission of this 3-wave-mixed field from the inside to the outside of the sample. Various factors of $\omega$ come from the Josephson relations between the phases and corresponding electric fields. Equipped with Eq.~\eqref{eqn:chi_3_gen}, we move on to analyse 2D THz spectroscopy experiments.
\section{2DTS of collective modes}\label{section_3}
This section is dedicated to developing an intuitive theoretical understanding of 2DTS in the context of collective excitations, in particular Josephson plasmons in layered superconductors. In Sec.~\ref{subssec:basics}, we describe the typical 2DTS protocol, whereby a sequence of terahertz pulses is sent onto the sample and a wave-mixing signal emitted either in the reflected or transmitted directions is measured (Figs.~\ref{fig:1}(a,b)). We discuss how such a protocol enables one to probe salient features of the third-order susceptibility. We also point out additional 2DTS considerations one must take into account when studying many-body systems. Subsequently in Secs.~\ref{subsec:rephasing} and~\ref{sec:mean_field_temp}, we clarify the applicability and shortcomings of the mean-field description developed in the preceding section when dealing with real condensed matter systems. Comparisons to the single-mode theory are drawn, which should guide the use of conventional intuition developed from atomic and molecular systems to collective modes in solids.

\subsection{Basic considerations of 2DTS in the many-body context}
\label{subssec:basics}
\subsubsection{ 2DTS protocol}

\begin{figure}[t!] 
\centering
\includegraphics[width=\linewidth]{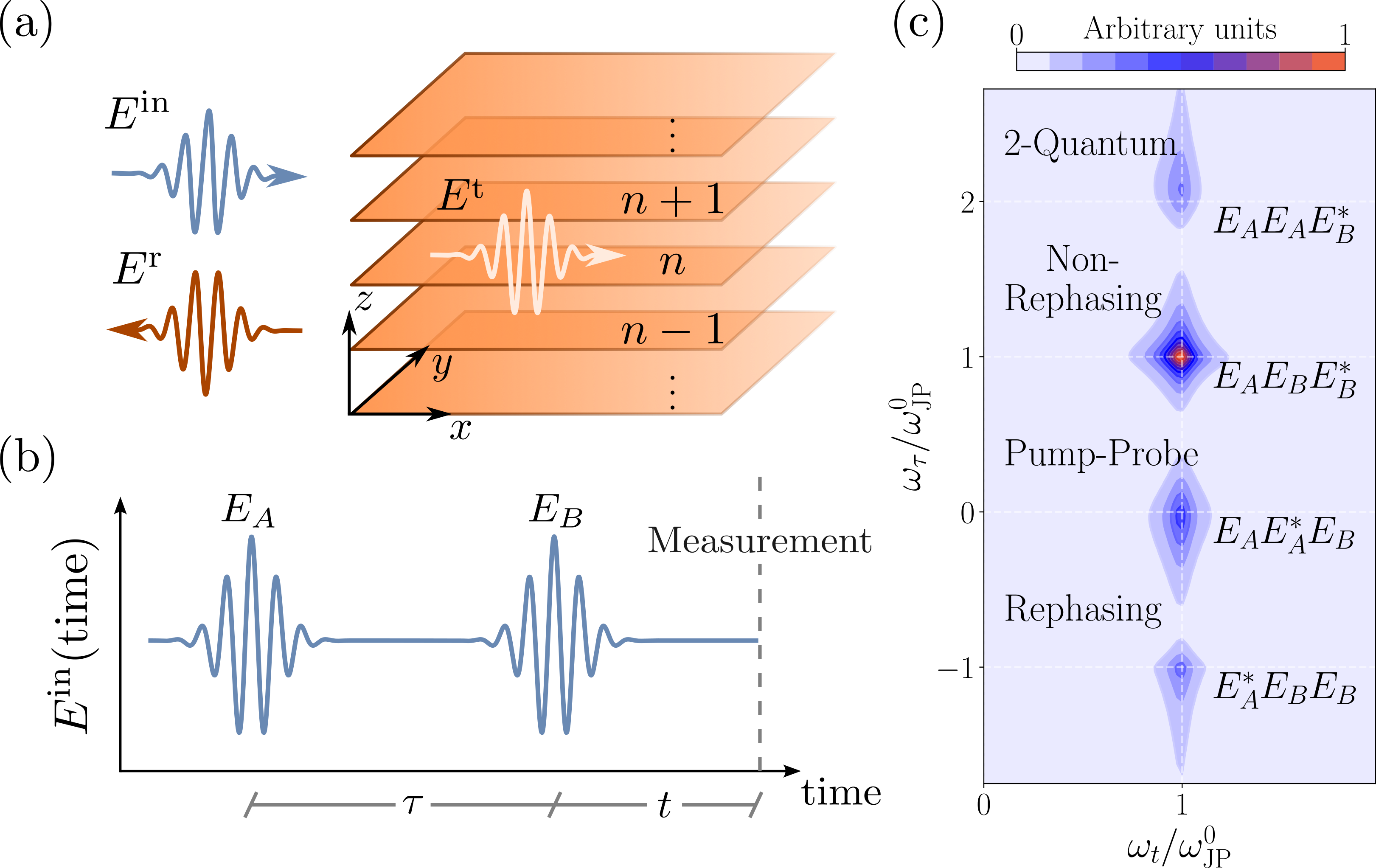}
\caption{Schematic of the setup. (a) 
Terahertz light $E^{\rm in}$ is sent onto the layered superconductor -- in 2D spectroscopy experiments specifically, one uses two pulses $E_A$ and $E_B$ separated from each other by time $\tau$ (b).
The resulting reflected radiation $E^{\rm r}$ is then measured at time $t$ after the arrival of the second pulse. (c) Typical 2D map. Up to third-order nonlinearities, it exhibits four distinctive peaks, each of which contains important information about Josephson plasma interactions. }
\label{fig:1}
\end{figure}

We first discuss the 2DTS protocol applied to a layered superconductor (Fig.~\ref{fig:1}(a)). We assume the incident field $E^{\rm in}$ is polarized along the out-of-plane $z$-direction and propagates along the in-plane $x$-direction. Resultant reflected and transmitted fields $E^{\rm r}$ and $E^{\rm t}$ are likewise polarized along the $z$-axis. The incident field consists of two identical excitation pulses, denoted $E_A$ and $E_B$, separated by a time delay $\tau$ as illustrated in Fig.~\ref{fig:1}(b). More precisely, the total incident electric field at the sample surface at the measurement time $t+\tau$ can be written as (zero of time is set to be the arrival of the first $A$-pulse):
\begin{equation}
    E^{\rm in}(x=0,t + \tau;\tau) = E_A(t + \tau)+E_B(t).
    \label{eqn: pulse}
\end{equation}

After interaction with both excitation pulses, the system is left to evolve unperturbed along time $t$, during which a nonlinear electric field $E_{\rm nl}$ (either reflected or transmitted) is measured (Fig.~\ref{fig:1}(b)). This nonlinear electric field emission is also measured as a function of the inter-pulse delay $\tau$, with sampling density and range parameters chosen according to the frequencies and linewidths of interest, and a two-dimensional array of measured values $E_{\rm nl}(t + \tau;\tau)$ is obtained. 
A Fourier transform with respect to both $\tau$ and $t$ defined as
\begin{equation}
    E_{\rm nl}(\omega_t ,\omega_\tau) = \int_0^\infty dt \int_0^\infty d\tau\, E_{\rm nl}(t+\tau;\tau) \, e^{i\omega_t t + i\omega_\tau \tau} \label{eqn:E_nl_2Dmap_FT}
\end{equation}
returns a two-dimensional spectrum $E_{\rm nl}(\omega_\tau,\omega_t)$, sometimes referred to as a `2D map'. Note that the integration in Eq.~\eqref{eqn:E_nl_2Dmap_FT} starts at $t = \tau = 0$ -- such definition ensures the causal relation $\tau > 0$ so that the $A$-pulse arrives first. A schematic (absolute-value) 2D map for a third-order nonlinearity of the Josephson plasmon is plotted in Fig.~\ref{fig:1}(c). As implied by the labels in Fig.~\ref{fig:1}(c), linear signals proportional to $E_A$ and $E_B$ as well as self-nonlinearities proportional to $E_A^3$ and $E_B^3$ are typically filtered out in experiments, leaving behind the mixing terms proportional to $E_A^2 E_B$ and $E_A E_B^2$. These terms are related to the third-order susceptibility as:
\begin{widetext}
\begin{align}
    E_{AAB}^{(3)}(\omega_t,\omega_\tau) &= 3\lim_{\substack{\delta_\tau\rightarrow0\\\delta_t\rightarrow 0}}\int\frac{d\omega_1}{2\pi}\int\frac{d\omega_2}{2\pi}\int\frac{d\omega_3}{2\pi}\,\chi^{(3)}(\omega_1,\omega_2,\omega_3)\,\frac{E_A(\omega_1)E_A(\omega_2)}{i(\omega_\tau-\omega_1-\omega_2)-\delta_\tau}\frac{E_B(\omega_3)}{i(\omega_t-\omega_1-\omega_2-\omega_3)-\delta_t }, \label{eqn:map_AAB} \\
    E_{ABB}^{(3)}(\omega_t,\omega_\tau) &= 3\lim_{\substack{\delta_\tau\rightarrow0\\\delta_t\rightarrow 0}}\int\frac{d\omega_1}{2\pi}\int\frac{d\omega_2}{2\pi}\int\frac{d\omega_3}{2\pi}\,\chi^{(3)}(\omega_1,\omega_2,\omega_3)\frac{E_A(\omega_1)}{i(\omega_\tau-\omega_1)-\delta_\tau}\,\frac{E_B(\omega_2)E_B(\omega_3)}{i(\omega_t-\omega_1-\omega_2-\omega_3)-\delta_t }.
    \label{eqn:map_ABB}
\end{align}
\end{widetext}
The relations~\eqref{eqn:map_AAB} and~\eqref{eqn:map_ABB} are generic and represent the starting point of our subsequent analyses. In the remainder of this section we discuss additional considerations that arise when transitioning from the conventional single-mode picture to a many-body system. Here we primarily focus on the mean-field nonlinearities encoded in Eq.~\eqref{eqn:chi_3_gen} and consider fluctuation corrections to $\chi^{(3)}$ in Sec.~\ref{sec:squeeze}.

\subsubsection{Pulse frequency filtering}

As follows from Eqs.~\eqref{eqn:map_AAB} and~\eqref{eqn:map_ABB}, the excitation spectra $E_A(\omega)$ and $E_B(\omega)$ act to filter out particular components of the third-order susceptibility. For conventional multidimensional spectroscopies of single-mode systems, the excitation pulse spectrum is typically far broader than any characteristic frequencies of the probed optical response. This permits approximating each excitation pulse as the idealized $\delta(t)$-function that directly samples the full nonlinear optical response function. This condition is therefore referred to as the `impulsive limit'.

In 2DTS experiments this condition is less well-defined, as for many-body systems one typically must  take into account a continuum of collective excitations, each of which can also have its own lifetime. 
For the mean-field case considered in this section, finite pulse effects are of no real utility. However, the filtering property of such finite pulses may become useful when physics beyond mean-field become prominent, which will be discussed in the following section (see Fig.~\ref{fig:6}).

\subsubsection{Effects due to the Boundary}
Another crucial difference between single-mode and many-body systems is the environment-sample interface. An immediate consequence of this interface is that only the magnitude and in-plane component (parallel to the interface) of momentum are conserved in the nonlinear wave-mixing processes. This consideration, due to symmetry breaking by the interface, is not important in the co-linear geometry considered here, but is crucial to determining emission wavevectors in a non-collinear excitation geometry \cite{Liu_2023_echo}. An additional consequence of the interface follows from Eq.~\eqref{eqn:chi_3_gen}, with the incoming (and outgoing) electric fields being further weighted by a transmission coefficient $t(\omega)$. As such, $t(\omega)$ acts as an additional `many-body filtering' on top of the aforementioned pulse frequency filtering. 
Let us illustrate this important interplay between the pulse-excitation and optical properties of the sample using the mean-field description of layered superconductivity developed above. For the one-dimensional polaritonic mode encoded in  Eq.~\eqref{eqn:1d_kx}, the dielectric function is given by:
\begin{equation}
    \epsilon(\omega;T) = \epsilon_\infty\Big( 1-\frac{(\omega_{\rm JP}(T))^2}{\omega(\omega+i0^+)}-\frac{\gamma(T)}{i\omega} \Big),
    \label{eqn:diell_func}
\end{equation}
where $\omega_{\rm JP}(T) = \sqrt{\Lambda_0(T)}$ is the Josephson plasmon resonance frequency and $\gamma(T)$ is the intrinsic decay rate, and both quantities generally depend on temperature $T$. For future reference, the reflection coefficient $R(\omega)$ and loss function $\mathcal{L}(\omega)$ are related to $\epsilon(\omega)$ through:
\begin{equation}
    R(\omega)=\abs{\frac{1-\sqrt{\epsilon(\omega)}}{1+\sqrt{\epsilon(\omega)}}}^2,\qquad  \mathcal{L}(\omega)=-\Im\{ \epsilon^{-1}(\omega) \}. 
    \label{eqn:lin_refl}
\end{equation}
It is worth pointing out that Eq.~\eqref{eqn:diell_func} could be used for fitting linear spectroscopy measurements to extract $\omega_{\rm JP}$ and $\gamma$. In fact, one of the appeals of the Bulayevskii framework we use here is that the fitting form~\eqref{eqn:diell_func} accurately captures actual experimental data~\cite{VANDERMAREL19911,Tamasaku_1992,koshelev1999fluctuation,Kuzmenko_2003}. When determining the filtering properties of the mean-field transmission coefficient $t(\omega)$, two considerations are relevant -- the density of states and dispersion of the polaritonic mode~\eqref{eqn:1d_kx} and its decay rate $\gamma$. At low temperatures, where the latter can be disregarded $\gamma \ll \omega_{\rm JP}$, $t(\omega)$ is sharply peaked at ${\omega}_{\rm JP}$, which, in particular, implies that i) the pulse frequency profile should be carefully chosen to have an appreciable spectral overlap with $t(\omega)$ and ii) the relevant probed frequencies are centered at $\omega_{\rm JP}$. At high temperatures, we instead expect $\gamma \gtrsim \omega_{\rm JP}$ and, as such, a featureless $t(\omega)$ such that the many-body filtering due to the interface is no longer that important. We further discuss this picture in Sec.~\ref{sec:mean_field_temp}.

\subsubsection{ Typical 2D maps}

We turn to briefly describe a typical 2D map of an anharmonic (classical or quantum) oscillator~\cite{friebolin1991basic,Mukamel2009,hamm_zanni_2011,book_MDCS}, which well captures the low-temperature behavior of Josephson plasmon nonlinearities. Throughout the remainder of the paper, we set $\omega_{\rm JP}^0 = \omega_{\rm JP}(T=0)$ to be the unit of energy.
Since $E(t)$ is a real variable, one gets $E(-\omega) = [E(\omega)]^*$ -- for this reason, we will discuss positive frequencies only and use complex conjugates for negative frequencies.

Equations~\eqref{eqn:map_AAB} and \eqref{eqn:map_ABB} give rise to four nonlinearities that radiate out at the fundamental Josephson plasma frequency $\omega_{\rm JP}^0$. Each of them corresponds to a unique combination of electric field interactions and, as such, appears as a distinct peak in the 2D map (see Fig.~\ref{fig:1}(c)). Each peak in Fig.~\ref{fig:1}(c) is labeled by its corresponding field interactions, which determine its position in the frequency space by multiplying the on-resonance phase factors $E_A \propto e^{-i\omega^0_{\rm JP}(t + \tau)}$ and $E_B \propto e^{-i\omega^0_{\rm JP} t}$. 

\begin{figure*}[t!]
    \centering
    \includegraphics[width=\linewidth]{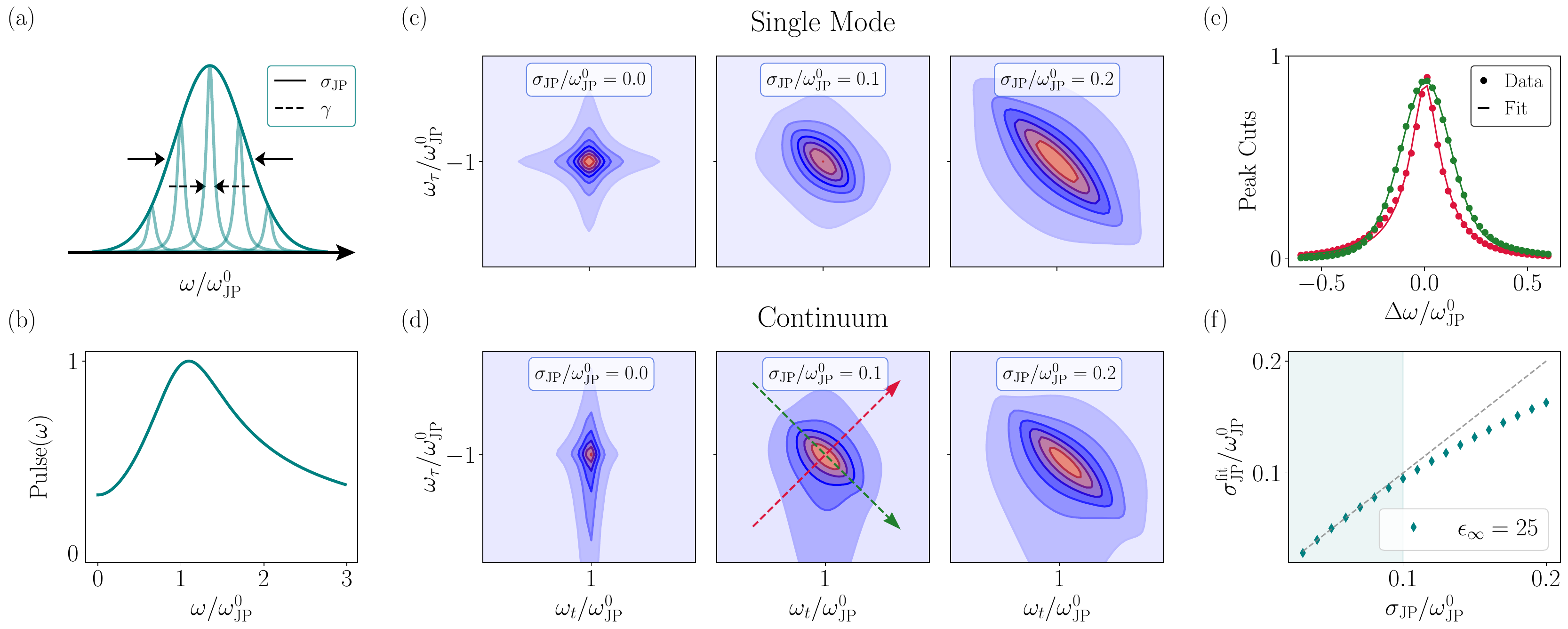}
    \caption{ The rephasing nonlinearity. (a) Sketch of the homogeneous (intrinsic) $\gamma$ and inhomogeneous  $\sigma_{\rm JP}$ broadenings. The latter is modeled as a static Gaussian distribution of the JP resonance, while the plasmon lifetime is kept fixed $\gamma/\omega_{\rm JP}^0=0.05$.
    (b) Pulse in Eq.~\eqref{eqn: Lorentzian Pulse} is chosen to have $\omega_d/\omega_{\rm JP}^0 = 1$ and $\sigma/\omega_{\rm JP}^0 = 0.5$.
   (c)-(d) Evolution of the rephasing peak with
   increasing inhomogenous disorder $\sigma_{\rm JP}$ for (c) single mode and (d) continuum of collective modes. In both cases, we observe that the star-shaped peaks turn into almond-shaped ones.
   (e) Slices across the diagonal (red) and cross-diagonal (green) lines of the continuum rephasing nonlinearity for $\sigma_{\rm JP}/{\omega}^0_{\rm JP} = 0.1$. Simultaneous fit of these two slices using single-mode functional forms of Ref.~\cite{Siemens_2010} enables one to estimate homogeneous and inhomogeneous broadenings: $\gamma^{\rm fit}/\omega_{\rm JP}^0 = 0.05 $, and $\sigma^{\rm fit}_{\rm JP}/\omega_{\rm JP}^0 = 0.098 $. (f) Such fitting is found to be reliable so long as $\sigma_{\rm JP}/\omega_{\rm JP}^0\lesssim 0.1$ (shaded area).}
    \label{fig:2}
\end{figure*}

We first address the peaks labeled `2-quantum' and `pump-probe'. One may intuitively understand the origin of each feature by first examining their interaction with the first excitation pulse $E_A$, which determines their position along the vertical frequency axis. From the perspective of the classical sine-Gordon nonlinearity, the 2-quantum and pump-probe peaks arise from parametric modulation (scaling with $E_AE_A$) and rectification (scaling with $E_AE_A^*$) of the resonance frequency, respectively. We point out that in the quantum oscillator picture (with initial state being vacuum), the 2-quantum peak arises from a coherence between the ground state and second excited state generated by the two field interactions $E_AE_A$, while the pump-probe peak arises from populations in either the ground or first excited state generated by $E_AE_A^*$.

In contrast, the peaks labeled `non-rephasing' and `rephasing' involve only a single interaction with the first excitation pulse --  with $E_A$ or $E_A^*$, respectively. From the classical perspective such interaction simply corresponds to displacing the Josephson plasmon coordinate, while from the quantum perspective it corresponds to generating a coherence between the ground state and first excited state. The difference between the two nonlinearities could be understood in terms of the relative phase of oscillations induced by the A-pulse to those of the subsequent emission induced by the B-pulse.  In the non-rephasing case, since the system oscillates with the same frequency $\omega_{\rm JP}$ after interacting with both pulses, this phase accumulates following each excitation. 

For the rephasing nonlinearity, the interaction with the first pulse gives rise to oscillations during time $\tau$ at $-\omega^0_{\rm JP}$. After two interactions with the second pulse, the state is then brought into a time-reversed superposition, which oscillates during time $t$ at $\omega^0_{\rm JP}$, implying instead the cancellation of the relative phase. This change of the frequency sign also implies that the initial state is restored whenever $\tau=t$, known as the celebrated `echo' phenomenon, which we turn to discuss in the many-body setting.

\subsection{The rephasing nonlinearity}
\label{subsec:rephasing}

The utility of rephasing `echoes' comes from the ability to disentangle energy disorder, as commonly performed with spin echoes in NMR and photon echoes in optical four-wave mixing~\cite{Hahn_1950,Siemens_2010,Bristow2011} of atomic and molecular systems. For a single-mode representation of the Josephson plasma resonance, such static energy disorder may be understood as $\omega_{\rm JP}^0$  is spanning a range of resonance frequencies (see Fig.~\ref{fig:2}(a)), introducing an inhomogeneous linewidth $\sigma_{\rm JP}$ in addition to the linewidth $\gamma$ from intrinsic level broadening. 

For collective excitations, the duration of terahertz optical pulses may become considerable with respect to the timescale of dynamics involved. For concreteness, from now on we model excitation pulses via:
\begin{equation}
    E_{A,B}(t) = \Theta(t) E_{A,B} e^{-\sigma t }\cos{\omega_d t},
    \label{eqn: Lorentzian Pulse}
\end{equation}
where $\omega_d$ is the pulse carrier frequency and $\sigma$ defines the spectral bandwidth of the excitation pulses (Fig.~\ref{fig:2}(b)). The choice of the form~\eqref{eqn: Lorentzian Pulse} is natural for two reasons: i) in the frequency domain, these pulses become simple Lorentzians, which facilitates analytical and numerical analyses, and ii) essentially any realistic pulse shape can be represented as a sum of Lorentzians.

The single-mode simulations presented in Figure~\ref{fig:2}(c) reproduces the conventional wisdom: As inhomogenous broadening $\sigma_{\rm JP}$ is tuned, the rephasing peak evolves from having a symmetric `star' shape when the intrinsic broadening dominates $\gamma \gg \sigma_{\rm JP}$ to being `almond'-shaped when $\sigma_{\rm JP}$ becomes appreciable. The values of $\gamma$ and $\sigma_{\rm JP}$ can then be directly obtained by simultaneously fitting the linecuts along the `diagonal' ($\omega_\tau = \omega_t$) and perpendicular `cross diagonal' directions of the $(1,-1)$-peak, using the fitting forms of Refs.~\cite{Siemens_2010,Bristow2011}.

\begin{figure}[t!]
    \centering
    \includegraphics[width=\linewidth]{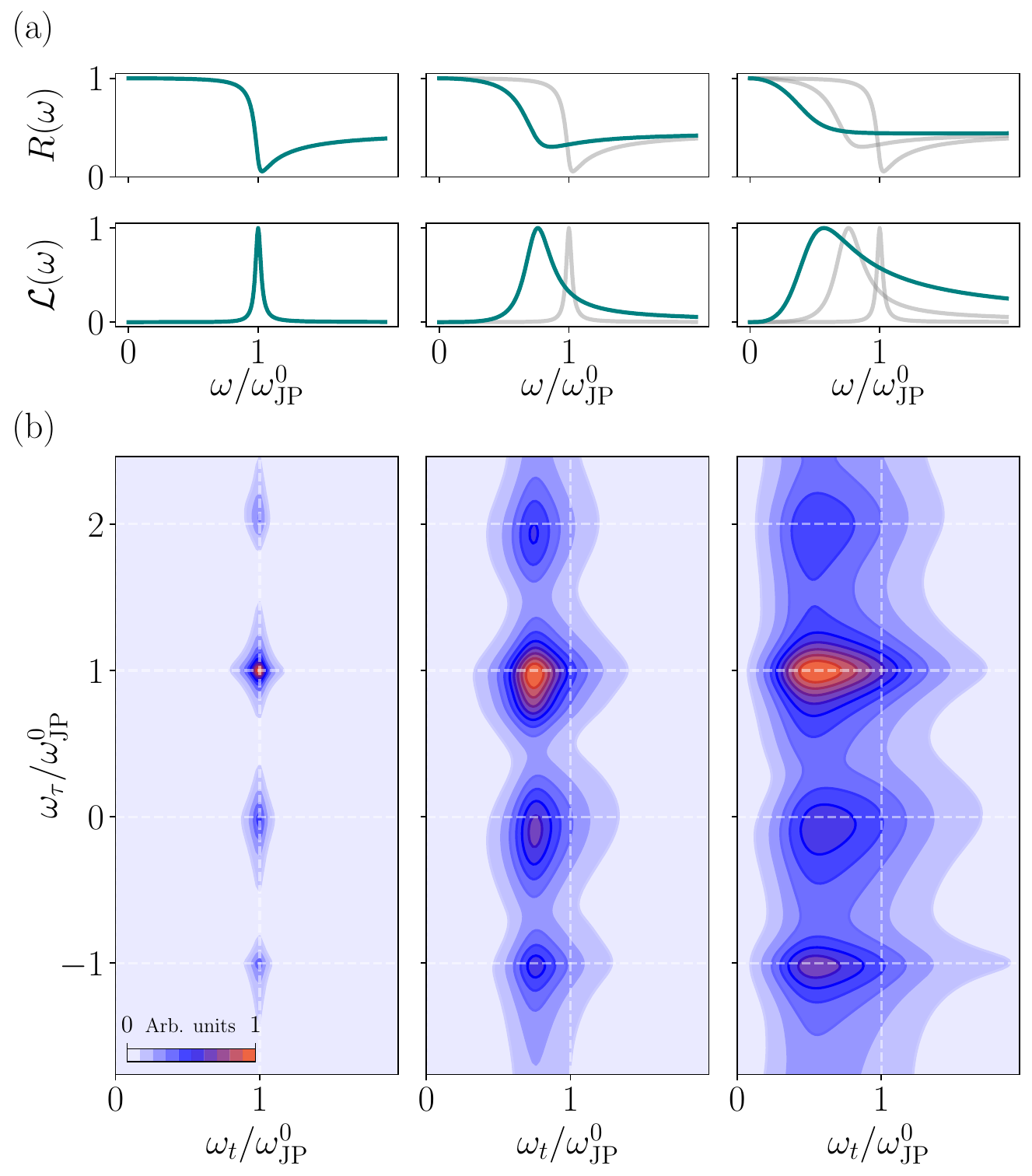}
    \caption{Evolution of 2D maps with temperature. (a) As $T$ increases, the JP resonance $\omega_{\rm JP}(T)$ softens and the decay $\gamma(T)$ grows, as can be deciphered from the linear reflectivity $R(\omega)$ and loss function $\mathcal{L}(\omega)$, Eq.~\eqref{eqn:lin_refl}. 
    (b) This behavior manifests in the mean-field 2D maps as horizontal sliding of the four peaks as well as their broadening. Here we fixed $\sigma/\omega_{\rm JP}^0 = 0.15$; the chosen temperatures  correspond to $\omega_{\rm JP}/\omega_{\rm JP}^0=\{ 1,0.75,0.5 \}$ and $\gamma/\omega_{\rm JP}^0=\{ 0.05,0.25,0.5 \}$, respectively.}
    \label{fig:3}
\end{figure}

This unique capability of the rephasing nonlinearity is clearly relevant to studying a wide range of quantum materials for which disorder plays an overt role in their properties. We remark that a reasonable starting point to understand how disorder affects Josephson plasmons is to consider static spatial inhomogeneities in the superfluid density -- this picture provided a clear interpretation to recent measurements of the rephasing nonlinearity in  La$_{2-x}$Sr$_x$CuO$_4$ (LSCO)~\cite{Liu_2023_echo}. However, how such disorder manifests in optical nonlinearities is a challenging many-body problem that might require developments based on the Keldysh and disorder-averaging techniques -- a task left for future work. Here we instead consider a static distribution of resonant frequencies $\omega_{\rm JP}^0$ in anticipation that, as it is the case for atomic and molecular systems, such modeling will prove useful to interpret future 2DTS experiments. Nevertheless, we still get an interesting nontrivial interplay between $\sigma_{\rm JP}$, $\gamma$, and the polariton dispersion~\eqref{eqn:1d_kx}, which makes our analysis of the many-body problem fairly distinct from the single-mode picture.

Quite strikingly, we find that, in the full continuum simulations presented in Fig.~\ref{fig:2}(d), the many-body rephasing peak essentially follows the single-mode phenomenology in Fig.~\ref{fig:2}(c). To quantitatively examine the agreement between these two scenarios, we use the single-mode fitting forms~\cite{Siemens_2010,Bristow2011} to extract $\gamma$ and $\sigma_{\rm JP}$ from the continuum 2D maps. Figure~\ref{fig:2}(e) shows the simulated lineshapes (taken along the corresponding arrows in Fig.~\ref{fig:2}(d)), and we find remarkable accuracy of such fitting. This agreement of the fitted disorder linewidth $\sigma^{\text{fit}}_{\text{JP}}$ is further examined quantitatively in Fig.~\ref{fig:2}(f), which returns the true linewidth to great accuracy for values of the disorder up to $\sigma_{\text{JP}}/\omega^0_{\text{JP}} \approx 0.1$, above which the disorder linewidth approaches the excitation bandwidth and fitted values begin to underestimate the disorder linewidth. In addition, we mention that our simplified modeling here well-captures both quantitative and qualitative aspects, including the peak shape as well as the extracted values of linewidths $\gamma$ and $\sigma_{\rm JP}$, of the reported rephasing nonlinearity of LSCO~\cite{Liu_2023_echo}.

\subsection{Approaching the phase transition}
\label{sec:mean_field_temp}

An interesting question we turn to address is the evolution of 2D maps with increasing temperature towards $T_c$. We will argue that the many-body filtering mentioned in Sec.~\ref{subssec:basics} will explicitly manifest, which is particularly appealing for the experimental verification of our predictions.

As $T$ is increased from $T = 0$ up to $T_c$, the initially sharp Josephson plasma resonance with $\gamma \ll \omega_{\rm JP}^0$ not only softens but also significantly broadens so that $\gamma \gtrsim \omega_{\rm JP}$ for $T \approx T_c$ -- this behavior is shown in Fig.~\ref{fig:3}(a), where the zero-temperature reflectivity plasma edge eventually becomes featureless, in qualitative agreement with the reflectivity experiments in LSCO~\cite{Tamasaku_1992,Kuzmenko_2003}.

The corresponding 2D maps, where the pulse spectra are chosen such as to have a substantial spectral overlap with the loss function~\cite{albert_phase_transition}, are shown in Fig.~\ref{fig:3}(b). Along the vertical frequency axis $\omega_\tau$, each peak is pinned by the peak frequency of the excitation spectrum and remains at the same position irregardless of temperature. Along the horizontal emission frequency axis $\omega_t$, however, the spectral weight of the peaks directly follow the linear response loss function due to the filtering property of the environment-sample interface. Let us remark that in the single-mode picture, where one is in the impulsive limit and uses identical excitation pulse spectra, one gets peaks arranged in the strict pattern shown in Fig.~\ref{fig:1}(c) so that these peaks would soften towards the origin $\omega_\tau = \omega_t = 0$ with decreasing plasma frequency.

We finally mention two recent 2DTS experiments on NbN~\cite{Katsumi2023_Arxiv} and LSCO~\cite{albert_phase_transition}. While NbN is not a layered superconductor, its nonlinearities are reasonably captured within mean-field theory, and, as such, the observations reported in Ref.~\cite{Katsumi2023_Arxiv} are found to be consistent with Fig.~\ref{fig:3}(b). Our theory also captures the low-temperature 2D maps of LSCO~\cite{albert_phase_transition}, but fails to explain the observations near $T \approx T_c$. Close to $T_c$, however, superconducting fluctuations become prominent, implying  that the mean-field description can break down -- we consider the role of plasma fluctuations in the following section and show that their correction to the third-order susceptibility explains the data of Ref.~\cite{albert_phase_transition} near $T_c$.

\section{Dynamical electromagnetic background}
\label{sec:squeeze}

The Josephson plasma resonance softening as $T$ approaches  $T_c$ is accompanied by proliferation of thermally excited plasmons, indicating that the above simplified modeling becomes insufficient. An intriguing possibility to explain the observed 2D maps of LSCO near $T_c$ could then be that the signal is dominated by the nonlinear process where the optical pulses give rise to a parametric drive of counter-propagating plasmon pairs of equal but opposite momenta -- a scenario we turn to investigate in this section. We remark that this process is at the heart of the plasmon squeezing proposal~\cite{dolgirev2022periodic} for light-induced superconductivity~\cite{von2022amplification}.

To account for the effects of dynamical electromagnetic background, we consider the Johnson-Nyquist normal-fluid noise~\cite{dolgirev2022periodic}, which modifies Eq.~\eqref{eqn:J_z_v0} to
\begin{align}
    J_{z;n,n+1} = J_0 \sin \varphi_{n,n+1} + \sigma_0 E_{z; n,n+1} + \xi^{}_{n}.
\end{align}
The fluctuation-dissipation theorem further imposes~\cite{kamenev2023field}:
\begin{align}
    \langle \xi_n(\bm r,t)\xi_m(\bm r',t')\rangle = \frac{2\sigma_0 T}{s}\delta_{nm} \delta(\bm r - \bm r') \delta(t-t').
\end{align}
Equation~\eqref{eqn: Bulaevskii} then acquires the Langevin form:
\begin{align}
     ( \partial_t^2 + \gamma \partial_t ) \psi_n - \nabla^2 L_{nm}\psi_m + \Lambda_0 \sin\psi_n = \xi_n^{\psi},
    \label{eqn: Bulaevskii noisy}
\end{align}
with
\begin{align}
    \langle\xi_n^{\psi}(\bm r,t) \xi_m^{\psi} (\bm r',t')\rangle = 2\gamma \tilde{T} \delta_{nm} \delta(\bm r - \bm r') \delta(t-t'),
\end{align}
where $\tilde{T} = 16 \pi e^2 s T/\varepsilon_{\infty} $. We note that Eq.~\eqref{eqn: Bulaevskii noisy} is derived under the assumption that the order parameter dynamics follows Eq.~\eqref{eqn:JR}, which might no longer hold near $T_c$ due to, for instance, proliferation of pancake-like vortices. However, even if one considers overdamped  order parameter dynamics (model-A in the classification of Ref.~\cite{hohenberg1977theory}), the Josephson plasma modes can remain long-lived excitations even above $T_c$~\cite{dolgirev2022periodic}. Additionally, the experimental analysis of the rephasing peak in optimally-doped LSCO~\cite{Liu_2023_echo} showed that the intrinsic broadening $\gamma$ dominates over disorder effects for $T\lesssim T_c$ (in Ref.~\cite{Liu_2023_echo}, $T\lesssim 0.7 T_c$). This suggests that pancake vortices are either nonessential or their effects can be well captured via a simple renormalization of the decay rate $\gamma(T)$, at least for temperatures  not too close to $T_c$. 
In the immediate vicinity of $T\approx T_c$, we expect that the Bulayevskii framework in Sec.~\ref{section_2} might become insufficient and should be revisited (we comment on this below).
We, thus, shall proceed with analysing Eq.~\eqref{eqn: Bulaevskii noisy}, as we expect that it should adequately describe the correct physics as $T$ approaches $T_c$.

\subsection{Gaussian fluctuations and equations of motion }
In the remainder of the paper, we treat the fluctuations within the Gaussian approximation and our derivations closely follow Refs.~\cite{PhysRevB.101.174306,dolgirev2022periodic}. In this section, we will also assume that external perturbations are homogeneous in space -- this will dramatically simplify our analysis, facilitating the understanding of the nonlinear physics due to the fluctuating electromagnetic background~\footnote{This assumption is partially justified near $T_c$, where the filtering effect due to the environment-superconductor interface, which is responsible for momenta mixing of the polaritonic mode~\eqref{eqn:1d_kx}, becomes unimportant because of the featureless transmission coefficient (see Sec.~\ref{section_3} and Fig.~\ref{fig:3}).}.
As a first step, we reduce the second-order differential equation~\eqref{eqn: Bulaevskii noisy} to coupled first-order ones by introducing the real-field $\pi_n(\bm r,t) = \partial_t \psi_n(\bm r,t)$:
\begin{gather}
    \partial_t \psi_n = \pi_n, \label{eqn:Bul_full_noise_psi}\\
    \partial_t \pi_n  + \gamma\pi_n - \nabla^2 L_{nm}\psi_m + \Lambda_0 \sin\psi_n = \xi_n^\psi.\label{eqn:Bul_full_noise_pi}
\end{gather}
The advantage of this simple reduction is that it allows us to promote the stochastic Langevin-like equations to the Fokker-Planck equation on the cumulative distribution function ${\cal P}[\psi_n(\bm r),\pi_n(\bm r);t]$. 
Within the Gaussian approximation, this time-dependent distribution function ${\cal P}[\psi_n(\bm r),\pi_n(\bm r);t]$ remains Gaussian even after photoexcitation. This, in particular, implies that the system's dynamics is fully characterized by the one- and two-point instantaneous correlation functions. 

The two one-point correlators, which are position independent due to translational invariance $\psi(t) = \langle \psi_n(\bm r,t) \rangle$ and $\pi(t) = \langle \pi_n(\bm r,t) \rangle$, satisfy:
\begin{gather}
    \partial_t \psi = \pi,\quad
    {\partial_t \pi  + \gamma\pi + \Lambda(t) \sin\psi = j_z(t),} \label{eqn:dyn_1p}
\end{gather}
where $j_z(t)$ encodes an external driving term assumed to be in the form of a charge current. The time-dependent coupling $\Lambda(t)$ is given by:
\begin{align}
    \Lambda(t) & \equiv \Lambda_0(t)\langle \cos \delta \psi_n(\bm r,t)\rangle  \notag\\
    & =\Lambda_0(t) \exp\Big[ -\frac{1}{2 {\cal A} N} \sum_{\bm q,q_z}{\cal D}_{\bm q,q_z}^{\psi\psi}(t)  \Big]. \label{eqn: Lambda}
\end{align}
The bare coupling $\Lambda_0(t)$ might be directly affected by the laser pulses, as we elaborate upon below. Here we have introduced the fluctuating field $\delta \psi_n(\bm r,t) = \psi_n(\bm r,t) - \psi(t)$ and its instantaneous two-point correlator ${\cal D}_{\bm q, q_z}^{\psi\psi}$:
\begin{align}
    {\cal D}_{\bm q, q_z}^{\psi\psi} (t) =\langle \delta\psi(-\bm q, -q_z; t) \delta\psi(\bm q, q_z; t) \rangle. \label{eqn:D_v0}
\end{align}
Interestingly, the time-dependent fluctuating background dynamically renormalizes the interaction strength $\Lambda_0 \to \Lambda(t)$ so that the effective model reminds the parametrically driven sine-Gordon.

Equations of motion for the two-point correlation functions, introduced as in Eq.~\eqref{eqn:D_v0}, read:
\begin{align}
    \partial_t {\cal D}^{\psi\psi}_{\bm q,q_z}  & =  2{\cal D}^{\psi\pi}_{\bm q,q_z} \label{eqn:dyn_2p_v1}
    ,\\
    \partial_t {\cal D}^{\psi\pi}_{\bm q,q_z}  & = {\cal D}^{\pi\pi}_{\bm q,q_z}  - \gamma {\cal D}^{\psi\pi}_{\bm q,q_z} \notag\\
    & 
    - [(q_x^2 + q_y^2) L(q_z) + \Lambda \cos \psi]{\cal D}^{\psi\psi}_{\bm q,q_z}\label{eqn:dyn_2p_v2} ,\\
    \partial_t {\cal D}^{\pi\pi}_{\bm q,q_z} & =
    2\gamma \tilde{T}  - 2\gamma{\cal D}^{\pi\pi}_{\bm q,q_z}\notag\\
    & 
    - 2 [(q_{x}^2 + q_y^2) L(q_z) + \Lambda\cos \psi ]{\cal D}^{\psi\pi}_{\bm q,q_z}. \label{eqn:dyn_2p_v3}
\end{align}
Here we have utilized the fact that all these correlation functions are real and ${\cal D}^{\pi\psi}_{\bm q,q_z} = {\cal D}^{\psi\pi}_{\bm q,q_z}$. 
\subsection{The plasmon squeezing mode and response functions}
We begin our analysis of the derived equations of motion~\eqref{eqn:dyn_1p},~\eqref{eqn:dyn_2p_v1}-\eqref{eqn:dyn_2p_v3} by examining the equilibrium correlation functions. In the absence of external time-dependent perturbations, we get $\psi = \pi = {\cal D}^{\psi\pi}_{\bm q,q_z} = 0$ and 
\begin{align}
    \bar{\cal D}^{\pi\pi}_{\bm q,q_z} = \tilde{T},\quad \bar{\cal D}^{\psi\psi}_{\bm q,q_z} = \frac{\tilde{T}}{(q_{x}^2 + q_y^2) L(q_z) + \Lambda_{\rm eq}   }, \label{eqn: eql}
\end{align}
where the equilibrium coupling $\Lambda_{\rm eq}$ is to be determined self-consistently:
\begin{align}
    \Lambda_{\rm eq} = \Lambda_0\exp\Big[ -\frac{1}{2 {\cal A} N} \sum_{\bm q,q_z}\bar{\cal D}_{\bm q,q_z}^{\psi\psi}\Big],
    \label{eqn: Lambda_eq}
\end{align}
We find that the Josephson plasma resonance $\omega_{\rm JP}^2(T) = \Lambda_{\rm eq}(T)$ softens with increasing $T$, even if we assume that the bare model parameters are temperature independent.
In fact, within the Gaussian approximation we employ here, the transition temperature at which the Josephson coupling is fully suppressed by fluctuations is found to be -- see Appendix~\ref{app: equilbirum}:
\begin{equation}
    \tilde{T}_c =\frac{8\pi c^2 s^2/\lambda_{ab}^2}{\epsilon_\infty\left(2+ s^2/\lambda_{ab}^2\right)}.\label{eqn: Tc}
\end{equation}
We remark, however, that this result does not take into account the fact that the Bulayevskii framework in Sec.~\ref{section_2} can become insufficient near the phase transition. In what follows, $\tilde{T}_c$ is used as the reference temperature.
As follows from Eq.~\eqref{eqn: eql}, which reflects the equipartition theorem, the softening can also be understood as proliferation of plasma fluctuations. For low temperatures, these plasma modes are barely populated and, thus, cannot manifest in response functions; as $\tilde{T}$ increases towards $\tilde{T}_c$, their role can no longer be ignored. 

We now turn to analyse collective excitations on top of the equilibrium state. Explicit linearization of the equations of motion~\eqref{eqn:dyn_1p},~\eqref{eqn:dyn_2p_v1}-\eqref{eqn:dyn_2p_v3} shows that the dynamics of one-point correlators $\psi$ and $\pi$ is decoupled from that of the ${\cal D}$-correlators. This can be understood as the fields $\psi$ and $\pi$ are IR-active since they directly couple to the external electric field drive and change sign under inversion; in contrast, the ${\cal D}$-correlators are Raman-active, remain intact under inversion, and, thus, cannot be excited via a single photon.

\begin{figure}
    \centering
    \includegraphics[width=\linewidth]{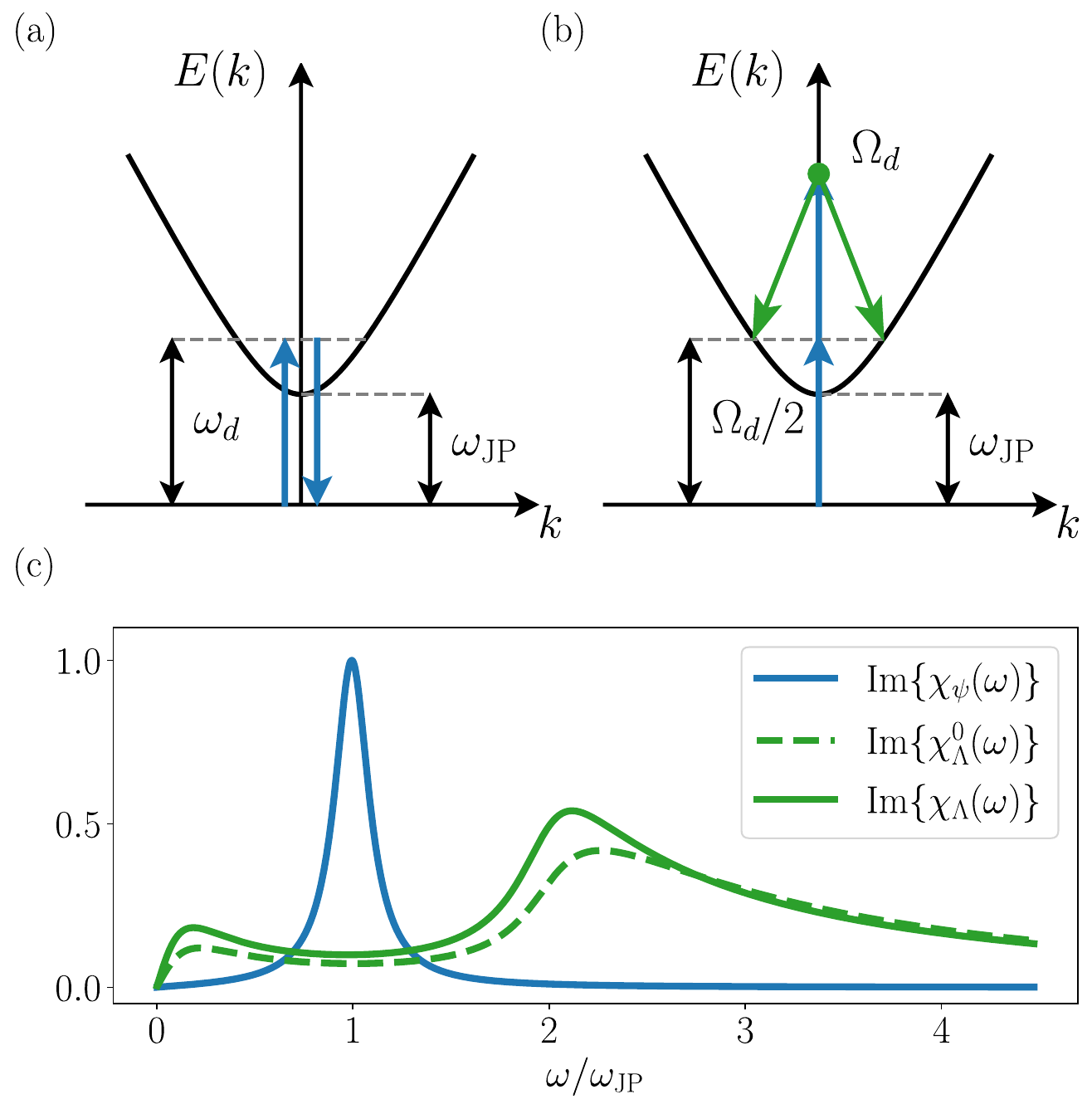}
    \caption{ Cartoon of the relevant (a) IR- (one-photon) and (b) Raman-like (two-photon) processes that determine optical properties of the sample.  The Raman drive at $\Omega_d$ excites a pair of counter-propagating plasmons, each of which has frequency $\Omega_d/2$.
    (c) One-photon $\chi_\psi(\omega)$  and two-photon $\chi_\Lambda(\omega)$ response functions. Most notably, we find that $\chi_\Lambda(\omega)$, Eq.~\eqref{eqn:chi_Lambda_v0}, is peaked in frequency at twice the Josephson plasma resonance $2\omega_{\rm JP}$. This is because the Raman process in (b) is amplified for $\Omega_d = 2\omega_{\rm JP}$, i.e., near the bottom of the plasmon band. Additionally, $\text{Im}[\chi_\Lambda(\omega)]$ is larger than $\text{Im}[\chi_\Lambda^0(\omega)]$ for $\omega\lesssim 2\omega_{\rm JP}$ -- an effect attributed to plasmon-plasmon interactions. All the responses are normalized by the maximum of $\text{Im}[\chi_\psi(\omega)]$.  
    }
    \label{fig:4}
\end{figure}

Not surprisingly, the linearized dynamics of $\psi$ and $\pi$ encodes nothing but the Josephson plasmons. To see this explicitly, one can compute the leading order response  $\psi^{(1)}(\omega) = \chi_\psi(\omega) j_z(\omega)$ to the $j_z(t)$-drive, cf. Eq.~\eqref{eqn:dyn_1p}:
\begin{align}
     \chi_\psi(\omega) = \frac{1}{-\omega^2 - i\gamma\omega + \Lambda_{\rm eq}} .
\end{align}
As in the preceding sections, this response function is sharply peaked in frequency at the Josepshon plasma resonance $\omega_{\rm JP}(T)$ -- see also Fig.~\ref{fig:4}. 

We turn to discuss the linearized dynamics of the two-point correlation functions. 
For reasons that will become more clear below and keeping in mind that an intense laser pulse might partially evaporate the superconducting condensate, we consider the following type of perturbations:
\begin{align}
    \Lambda_0 \to \Lambda_0 + \delta \Lambda_0(t).
    \label{eqn: delta_Lambda}
\end{align}
While such a perturbation of the bare coupling constant leaves the one-point correlators intact $\psi = \pi = 0$, it can result in a nontrivial dynamics of the two-point correlators 
$\mathcal{D}^{\alpha\beta}(t) = \bar{\mathcal{D}}^{\alpha\beta} + \delta\mathcal{D}^{\alpha\beta}(t)$ and, as such, of the renormalized coupling $\Lambda(t) = \Lambda_{\rm eq} + \delta\Lambda(t)$. To the leading order in $\delta \Lambda_0(t)$, we get
\begin{align}
    \delta\Lambda(t) & = \Lambda_{\rm eq} \Big[ 
     \frac{\delta \Lambda_0(t)}{\Lambda_0}  - \frac{1}{2 {\cal A} N} \sum_{\bm q,q_z}
     \delta{\cal D}_{\bm q,q_z}^{\psi\psi}(t)
    \Big]. \label{eqn:coupled_in}
\end{align}
This expression written in the frequency domain
\begin{align}
    \delta\Lambda(\omega) = [1 + \chi_\Lambda(\omega)] \delta \Lambda_0(\omega) \Lambda_{\rm eq}/\Lambda_{0}
\end{align}
allows us to introduce the response function $\chi_\Lambda(\omega)$, the central object of the upcoming discussion. Linearization of Eqs.~\eqref{eqn:dyn_2p_v1}-\eqref{eqn:dyn_2p_v3} gives:
\begin{gather}
   -i \omega \delta {\cal D}^{\psi\psi}_{\bm q,q_z} = 2 \delta {\cal D}^{\psi\pi}_{\bm q,q_z},\label{eqn:coupled_in_v2}\\
    -i(\omega + 2i\gamma)\delta{\cal D}^{\pi\pi}_{\bm q,q_z}  = -2 [
     \bm q^2 L(q_z) + \Lambda_{\rm eq} ]
     \delta{\cal D}^{\psi\pi}_{\bm q,q_z},
\end{gather}
and 
\begin{align}
    - i(\omega + i\gamma)\delta{\cal D}^{\psi\pi}_{\bm q,q_z}  = \delta{\cal D}^{\pi\pi}_{\bm q,q_z} -[ &
     \bm q^2 L(q_z)  + \Lambda_{\rm eq} ]
     \delta{\cal D}^{\psi\psi}_{\bm q,q_z} \notag\\
    &\qquad- \delta\Lambda(\omega)
     \bar{\cal D}^{\psi \psi}_{\bm q,q_z}. \label{eqn:coupled_fin}
\end{align}
Solving these coupled equations~\eqref{eqn:coupled_in}-\eqref{eqn:coupled_fin}, we obtain (see Appendix~\ref{Appendix: dynamics of correlators} for details):
\begin{equation}
    \chi_\Lambda (\omega) =  \frac{\chi^0_\Lambda (\omega)}{1 - \chi^0_\Lambda (\omega)}, 
    \label{eqn:chi_Lambda_v0}
\end{equation}
where
\begin{align}
    \chi^0_\Lambda (\omega)  = -& \frac{2\tilde{T}\Lambda_{\rm eq}}{\tilde{T}_c\omega(\omega+i\gamma)} \log\left( 1-i\frac{\gamma\omega}{2\Lambda_{\rm eq}}-\frac{\omega^2}{4\Lambda_{\rm eq}} \right).\label{eqn:chi_Lambda_v1}
\end{align}
While during Raman perturbations, such as in Eq.~\eqref{eqn: delta_Lambda}, the average electromagnetic field remains zero ($\psi = \pi = 0$), the dynamics of the average electromagnetic energy density, encoded in the ${\cal D}$-correlators, can be nontrivial. This is the reason the response functions $\chi_\Lambda^0(\omega)$ and $\chi_\Lambda(\omega)$ are associated with the Josephson plasmon squeezing (for additional discussion, see Ref.~\cite{dolgirev2022periodic}).

The response function $\chi_\Lambda^0(\omega)$ has a rather involved and interesting structure shown in Fig.~\ref{fig:4}(a). Most remarkably, we find that $\text{Im}[\chi_\Lambda^0(\omega)]$ is peaked in frequency at around $2\omega_{\rm JP}$. To understand this feature, we note that a Raman drive at frequency $\Omega_d$, as in Eq.~\eqref{eqn: delta_Lambda}, necessarily excites a pair of counter-propagating Josephson plasmons, each with a frequency of $\Omega_d/2$ (see Fig.~\ref{fig:4}(b)). Such a Raman process is enhanced when $\Omega_d/2$ matches the bottom of the plasmon band $\Omega_d = 2\omega_{\rm JP}$, i.e., where the density of states exhibits a van Hove singularity (see Fig.~\ref{fig:4}). 

Other interesting features of $\text{Im}[\chi_\Lambda^0(\omega)]$ include a hump at small frequencies $\omega\lesssim \omega_{\rm JP}$ and a slow $\sim 1/\omega^2$ decay at large frequencies $\omega\gtrsim 2\omega_{\rm JP}$. Both these effects originate from the interplay between the plasmon density of states, which grows with $\omega$ for $\omega \geq \omega_{\rm JP}$, and $\omega$-dependent matrix elements, associated with the plasmon propagation (see also Eq.~\eqref{eqn:JR}), that decrease with $\omega$, as can be inferred from the prefactor of Eq.~\eqref{eqn:chi_Lambda_v1}. Indeed, for small frequencies, while the plasmon pair generation appears highly off-resonant, the mentioned matrix elements diverge for $\omega \to 0$ so that the net effect manifests as the non-monotonic hump seen for $\omega\lesssim \omega_{\rm JP}$. Similarly, for large frequencies, these matrix elements decay faster than the growth of the density of states so that $\text{Im}[\chi_\Lambda^0(\omega)]$ decreases with $\omega$ for $\omega\gtrsim 2\omega_{\rm JP}$.

\begin{figure*}[t!]
    \centering
    \includegraphics[width=0.8\linewidth]{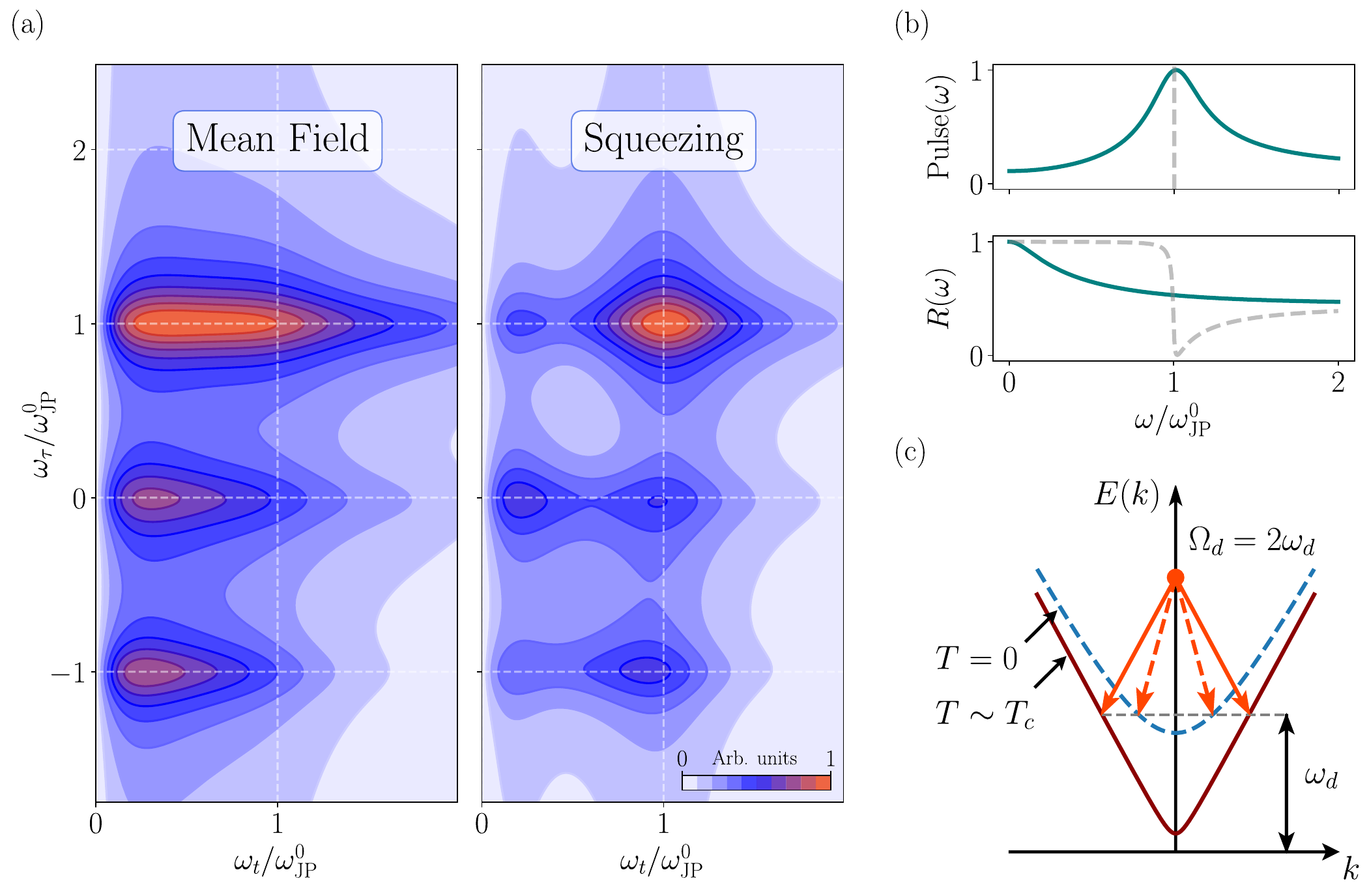}
    \caption{Josephson plasmon squeezing and 2D spectroscopy. (a) 2D maps of the single mode, Eq.~\eqref{eqn:chi_3_mf}, and squeezing channel, Eq.~\eqref{eqn:chi_3_sq}. The single mode 2D map (left) is in agreement with Fig.~\eqref{fig:3}, as it reveals an elongated shape along the $\omega_\tau$-axis. The squeezing 2D map (right) exhibits a resonance centered at $\omega_t \approx \omega_\tau \approx \omega_{\rm JP}^0$ (the (1,1)-peak), even when the plasma frequency $\omega_{\rm JP}$ has appreciably softened, as evidenced by the featureless plasma edge in the linear reflectivity (b). (c) Cartoon explaining the origin of this peaked behavior. We note that the  Raman driving (see text and Fig.~\ref{fig:4}) occurs at $\Omega_d = 2\omega_d$, where $\omega_d \approx \omega_{\rm JP}^0$ is the pulse frequency (b). While the plasmons soften as $T$ approaches $T_c$, their dispersion is barely affected near $\omega_{\rm JP}^0$, in turn explaining why the squeezing peak position is not sensitive to $T$. }
    \label{fig:5}
\end{figure*}

One can intuitively think of $\chi_\Lambda^0(\omega)$ as the response function when the plasmon-plasmon interactions are neglected and of $\chi_\Lambda(\omega)$ as it takes into account these interactions within the RPA (see also Appendix~\ref{Appendix: dynamics of correlators}). We find that the overall behavior of $\chi_\Lambda(\omega)$ is similar to $\chi^0_\Lambda(\omega)$, except $\text{Im}[\chi_\Lambda(\omega)]$ is a bit enhanced compared to $\text{Im}[\chi^0_\Lambda(\omega)]$ for $\omega\lesssim 2\omega_{\rm JP}$ (see Fig.~\ref{fig:4}(c)). This effect originates from the fact that the attractive plasmon-plasmon interactions can give rise to a bi-plasmon binding. However, for parameters relevant for cuprate superconductors, the binding energy is found to be negligibly small, manifesting only as the mentioned enhancement of $\text{Im}[\chi_\Lambda(\omega)]$. For this reason, we will further discuss 2DTS signatures of the bi-plasmon binding elsewhere. Most of our conclusions can be intuitively understood by neglecting the plasmon interactions.

\subsection{ Plasma fluctuations in 2DTS}
In this subsection, we argue that the fluctuating electromagnetic background, while negligible at low temperatures, dramatically affects the 2DTS signal as $T$ approaches $T_c$. In particular, such measurements give access to the squeezing response function $\chi_\Lambda(\omega)$, which is then shown to provide a natural interpretation of experimental 2D maps. Even more strikingly, the pulse filtering properties discussed in Sec.~\ref{section_3} enable us to unambiguously distinguish the non-mean-field squeezing response from the mean-field nonlinearities -- this, in turn, is argued to enable us to probe thermally excited finite-momentum plasmons.

To mimic 2DTS experiments, we evaluate the third-order response to
current perturbations as in Eq.~\eqref{eqn:dyn_1p}. Specifically, a train of light pulses, Eq.~\eqref{eqn: pulse}, is now modeled via
\begin{equation}
    j(t + \tau;\tau) = j_A(t + \tau)+j_B(t),
    \label{eqn: pulse_j}
\end{equation}
with
\begin{equation}
    j_{A,B}(t) = j_0 \Theta(t)  e^{-\sigma t }\cos{(\omega_d t)}.
\end{equation}

The third-order response can be split into two contributions:
\begin{align}
    \psi^{(3)}(t)  = \psi^{(3)}_{\rm mf}(t) + \psi^{(3)}_{\rm sq}(t).\label{eqn:psi_3_hom}
\end{align}
The first mean-field term $\psi^{(3)}_{\rm mf}$ is fully analogous to that discussed in Sections~\ref{section_2} and~\ref{section_3}. Furthermore, the corresponding nonlinear response function (see Appendix~\ref{appendix:3rd order})
\begin{align}
    \chi_{\rm mf}^{(3)}(\omega_1,\omega_2,\omega_3) = \frac{\Lambda_{\rm eq}}{6} & \chi_\psi(\omega_1 + \omega_2 + \omega_3)\notag\\
    &\quad \chi_\psi(\omega_1)\chi_\psi(\omega_2)\chi_\psi(\omega_3)
    \label{eqn:chi_3_mf}
\end{align}
can be written solely in terms of the linear response function $\chi_\psi(\omega)$, cf. Eq.~\eqref{eqn:chi_3_gen}. Since here we consider homogeneous perturbations, $\psi^{(3)}_{\rm mf}$ encodes only a single mode with no momenta mixing, naturally present in the preceding analysis due to the environment-superconductor interface. Nevertheless, we find that the resulting mean-field 2D map near $T_c$ qualitatively follows the discussion in Sec.~\ref{sec:mean_field_temp} because we use pulse spectra that are narrower than the intrinsic broadening (see Fig.~\ref{fig:3}(b)) -- as such, the term $\psi^{(3)}_{\rm mf}$ cannot explain the observed 2DTS measurements for $T\lesssim T_c$.
\begin{figure*}
    \centering
    \includegraphics[width=0.95\linewidth]{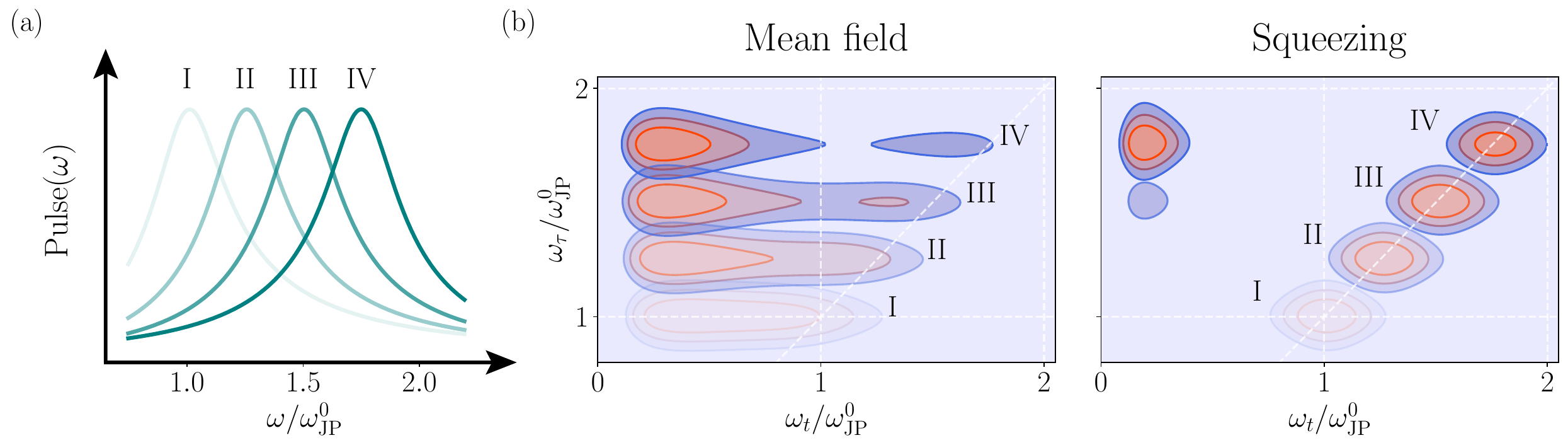}
    \caption{ Disentangling the mean-field and non-mean-field nonlinearities. (a) Sketch of a carrier frequency $\omega_d$ sweep for fixed pulse width $\sigma/\omega_{\rm JP}^0=0.15$. (b)  As $\omega_d$ is tuned, the mean-field nonlinearity drifts vertically following $\omega_\tau=\omega_d$ and $\omega_t=\omega_{\rm JP}$. In contrast, the squeezing nonlinearity drifts diagonally and is centered at $\omega_\tau=\omega_t=\omega_d$.}
    \label{fig:6}
\end{figure*}
The second term $\psi^{(3)}_{\rm sq}$ in Eq.~\eqref{eqn:psi_3_hom} is new and 
encodes the dynamics of a fluctuating electromagnetic background. The corresponding third-order response function is given by (see Appendix~\ref{appendix:3rd order}):
\begin{align}
   & \chi_{\rm sq}^{(3)}(\omega_1,\omega_2,\omega_3) = \frac{\Lambda_{\rm eq}}{6} \sum_{i = 1,2,3}\chi_\psi(\omega_1 + \omega_2 + \omega_3) \label{eqn:chi_3_sq}\\ 
   &\qquad\qquad \times\chi_\Lambda(\omega_1+\omega_2 + \omega_3 -\omega_i)\chi_\psi(\omega_1)\chi_\psi(\omega_2)\chi_\psi(\omega_3) . \notag
\end{align}
Most notably, $\chi_{\rm sq}^{(3)}$ gives access to the squeezing response function $\chi_\Lambda(\omega)$, which is beyond both the linear optical response and preceding mean-field modeling. The squeezing channel in Eq.~\eqref{eqn:chi_3_sq} becomes relevant for 2DTS only near $ T_c$, where the thermal population of the plasma modes is appreciable, cf. Eq.~\eqref{eqn: eql}.
\subsubsection{2D maps as {$T$} approaches {$T_c$}}
We now turn to discuss the unique signatures of the non-mean-field nonlinearity $\chi_{\rm sq}^{(3)}$ in 2DTS. In contrast to the mean-field 2D maps that feature a peak that softens following the linear loss function, the squeezing channel manifests as a peak centered at $\omega_t \approx \omega_\tau \approx \omega_d$ regardless of $T$ (Fig.~\ref{fig:5}(a)).  The squeezing 2D maps exhibit i) a dominant non-rephasing nonlinearity and ii) the corresponding peak positions set by the carrier frequency $\omega_d$ are independent of temperature -- these two features both agree with and explain recent 2DTS measurements  near $T\approx T_c$ on optimally-doped LSCO~\cite{albert_phase_transition}.

This peaked behavior can be physically understood as follows. We first note that close to $T_c$, the softening of the plasmon resonance $\omega_{\rm JP}$ is accompanied by a dramatic increase in its intrinsic decay rate. As such, for $\gamma \gg \omega_{\rm JP}$, reflectivity appears featureless (Fig.~\ref{fig:5}(b)).
Therefore, even off-resonant excitation at $\omega_d\gg\omega_{\rm JP}$ can launch a plasmon with the same frequency (more generically, this could be a virtual process) which, as we further elaborate upon in Appendix~\ref{appendix:3rd order}, can act as a Raman-like driving $\sim \psi^2(t)$ at frequency $\Omega_d = 2\omega_d$. From the perspective of the downconversion illustrated in Fig.~\ref{fig:4}, this Raman drive parametrically excites counter-propagating Josephson plasmons with frequency $\Omega_d/2 = \omega_d$ (this process can also be understood within Eq.~\eqref{eqn:dyn_1p}, where $\Lambda(t)$ oscillates at $\Omega_d$). The squeezing nonlinearity therefore leads to generating a continuum of plasmon pairs with the frequency distribution set by the excitation spectrum, which determines the peak position along the vertical axis $\omega_\tau \approx \omega_d$. The featureless reflectivity spectrum then leads to an emission spectrum that likewise follows the excitation spectrum, centered at $\omega_t \approx \omega_d$. We finally note that for $\omega_d \gg \omega_{\rm JP}$, the plasmon dispersion near $\omega_d$ is barely affected by $T$ (see Fig.~\ref{fig:5}(c)), further supporting the robustness of the peak position to temperature.
\begin{figure*}
    \centering
    \includegraphics[width=0.9\linewidth]{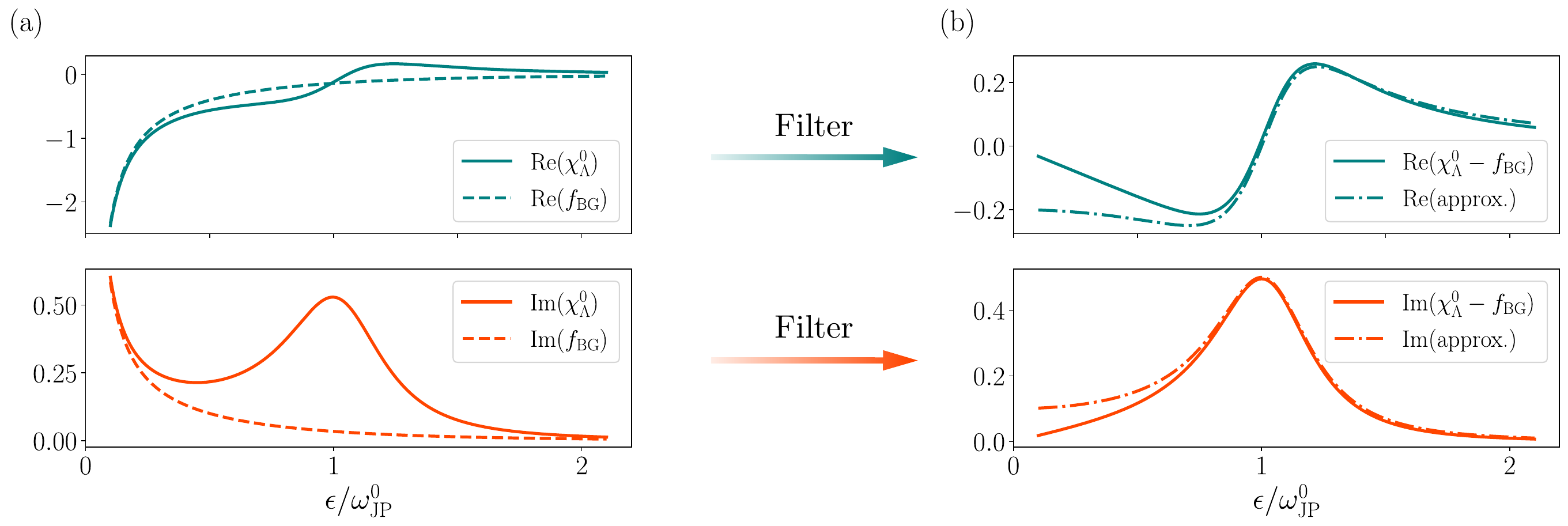}
    \caption{ Probing finite-momentum thermal fluctuations. (a) Real and imaginary parts of  the integrand of Eq.~\eqref{eqn:chi_0_integ} (solid) and the `background' function $f_{\rm BG}(\epsilon,\omega)$ in Eq.~\eqref{eqn_f_BG} (dashed). By subtracting one from the other (b), we get a signal that is well represented by a single  Lorentzian (dashed-dotted) that carries information about ${\cal N}(\omega_d) \times n_{\omega_d}$. Parameters used: $\gamma/\omega_{\rm JP}^0 = 0.5$,  $\omega_d/\omega_{\rm JP}^0 = 1$, and $\omega_{\rm JP}/\omega_{\rm JP}^0 = 0.1$.}
    \label{fig:fluctuations}
\end{figure*}
\subsubsection{Phenomenological model above $T_c$}

Extending our approach to temperatures above $T_c$ requires additional considerations. There is strong experimental evidence that in high-$T_c$ cuprates local superconducting correlations remain finite even when long-range order disappears~\cite{wang2006nernst,li2007two,Li_TorqueMagnet_2010,bilbro2011temporal,Grbic_MicrowaveAbs_2011,Zhou_ShotNoise_2019}.  Following Refs.~\cite{Koshelev,koshelev1999fluctuation} we expect that in this case one can separate slow statistical fluctuations $\psi_n^{\rm stat}$ of the phase of the  order parameter and fast fluctuations $\tilde{\psi}_n$ describing collective modes and response to electromagentic probes. Starting from Eq.~\eqref{eqn: Bulaevskii} 
\begin{gather}
(\partial_t^2 +\gamma \partial_t) (\psi^{\rm stat}_n + \tilde{\psi}_n) - \nabla^2 L_{nm}(\psi^{\rm stat}_m + \tilde{\psi}_m)+ \notag\\ {\Lambda}_0 \sin (\psi^{\rm stat}_n + \tilde{\psi}_n) =0
\label{Statistical_and_Plasmons},
\end{gather}
one is required to average over the statistical phase fluctuations $\psi^{\rm stat}_n$ to obtain an effective model for $\tilde{\psi}_n$, which describes collective modes and terahertz electromagnetic response. Plasmon type collective excitations are expected to persist above $T_c$, but they should become gapless and have stronger damping. We expect that effective theory of such plasmons can be captured by a model of the type:
\begin{eqnarray}
(\partial_t^2 +\tilde{\gamma} \partial_t) \tilde{\psi}_n - \nabla^2 L_{nm}\tilde{\psi}_m + \tilde{\Lambda}_4 \tilde{\psi}_n^3 =0
\label{Plasmons_above_Tc}.
\end{eqnarray}
Linearized form of Eq.~\eqref{Plasmons_above_Tc} gives a linearly dispersing mode at small momenta, and the last term describes nonlinear interaction between plasmons. 

It is useful to note that Eq.~\eqref{Plasmons_above_Tc} can not be obtained by simply setting ${\Lambda}_0 $ to zero in Eq.~\eqref{eqn: Bulaevskii}. While the latter naive approach gives a gapless plasmon spectrum, it results in vanishing nonlinear interaction between plasmons, Eqs.~\eqref{eqn:chi_3_mf} and~\eqref{eqn:chi_3_sq}. We postpone detailed discussion of plasmons in the pseudogap regime until future publication. 
\subsubsection{Disentangling the non-mean-field nonlinearity}

We note that for a wide variety of parameters the mean-field rephasing nonlinearity may obscure the squeezing contribution, as can be seen in Fig.~\ref{fig:5} and further expanded upon below in the context of third harmonic generation. An exciting possibility, rather unique to 2DTS, is to exploit the filtering effect of the excitation pulse spectrum discussed in Sec.~\ref{subssec:basics} in order to unambiguously disentangle the non-mean-field response $\chi_{\rm sq}^{(3)}$ from the  mean-field one $\chi_{\rm mf}^{(3)}$ (see Fig.~\ref{fig:6}). 
Indeed, following our discussion in Sec.~\ref{sec:mean_field_temp}, we expect that tuning the carrier frequency $\omega_d$ (Fig.~\ref{fig:6}(a)) will result in a vertical shift of the non-rephasing mean-field nonlinearity according to $\omega_\tau \approx \omega_d$ (Fig.~\ref{fig:6}(b)), provided the pulse frequency profile has appreciable spectral overlap with the loss function. At the same time, as follows from the preceding discussion, the squeezing peak is locked to $\omega_t \approx\omega_\tau \approx \omega_d$ and, thus, upon tuning $\omega_d$, will shift diagonally (Fig.~\ref{fig:6}(b)), thereby  separating the two responses from each other.

\subsubsection{Probing finite-momentum thermal  fluctuations}
This capability of 2DTS to disentangle the squeezing nonlinearity can then be used to extract useful information about thermally-excited plasmons at finite momentum. Intuitively one expects, and we demonstrate this below, that a careful analysis of the non-rephasing nonlinearity (see Fig.~\ref{fig:6}) gives access to ${\cal N}(\omega_d) \times n_{\omega_d}$, where ${\cal N}(\omega_d)$ is the plasmon density of states at $\omega_d$ and $n_{\omega_d} = T/\omega_d$ is the thermal occupation number of the plasmons with energy $\omega_d$. 

To this end, we rewrite $\chi_\Lambda^0(\omega)$ as (cf. Eq.~\eqref{eqn: chi_int}):
\begin{equation}
    \chi_\Lambda^0(\omega) = -\frac{\omega+2i\gamma}{\omega+i\gamma}\int_0^\infty d\epsilon\,\frac{\Lambda_{\rm eq}\,\mathcal{N}(\epsilon)\,\bar{\cal D}^{\psi\psi}_{\epsilon}}{\omega(\omega+2i\gamma)-(2\epsilon)^2}, 
    \label{eqn:chi_0_integ}
\end{equation}
where $\bar{\cal D}^{\psi\psi}_{\epsilon}=\tilde{ T}/\epsilon^2$, cf. Eq.~\eqref{eqn: eql}, and
\begin{equation}
    \mathcal{N}(\epsilon) = \Theta(\epsilon-\omega_{\rm JP})\underbrace{(2+s^2/\lambda_{ab}^2)\frac{\lambda_{ab}^2\epsilon_\infty}{ 2\pi c^2 s^2}}_{\equiv\,\nu} \,\epsilon.
\end{equation}
Following the preceding discussion (see also Eq.~\eqref{eqn:chi_3_sq} and Figs.~\ref{fig:5} and~\ref{fig:6}), we are interested in evaluating $\chi_\Lambda^0(\omega)$ at around $\omega=2\omega_d$ for $\omega_d \gg \omega_{\rm JP}$. We note that the denominator of the integrand in Eq.~\eqref{eqn:chi_0_integ} suggests that we get a Lorentzian peaked at $\epsilon=\omega_d$; however, the $1/\epsilon^2$-behavior of $\bar{\cal D}^{\psi\psi}_{\epsilon}$ gives rise to a more complicated shape of this integrand shown in Fig.~\ref{fig:fluctuations}(a) -- in fact, the contribution at small $\epsilon\approx\omega_{\rm JP}$ can obscure the peaked behavior near $\epsilon=\omega_d$ that we are interested in. The latter can be isolated via introducing the `background' function ($\gamma\lesssim\omega_d$)
\begin{equation}
    f_{\rm BG}(\epsilon,\omega) = \Lambda_{\rm eq}\nu T\frac{-\omega^2+i\gamma\omega}{4\epsilon^3(\omega^2+\gamma^2)}\tanh^2\left(\frac{2\epsilon}{\omega}\right) 
    \label{eqn_f_BG}
\end{equation}
that well captures the asymptotic behavior (both $\epsilon\to 0$ and $\epsilon\to \infty$) of the integrand of Eq.~\eqref{eqn:chi_0_integ}, as shown in Fig.~\ref{fig:fluctuations}(a). Assuming that both ${\cal N}(\epsilon)$ and $\bar{\cal D}^{\psi\psi}_\epsilon$ vary slowly near $\epsilon = \omega_d$ (see Fig.~\ref{fig:fluctuations}(b)), we thus arrive at:
\begin{align}
     \chi_\Lambda^0(2\omega_d) & \simeq \int_{\omega_{\rm JP} }^{\infty} \, f_{\rm BG}(\epsilon,2\omega_d) \, d\epsilon \,  \\ & +\Lambda_{\rm eq}\mathcal{N}(\omega_d)\bar{\cal D}^{\psi\psi}_{\omega_d} \int_0^\infty \frac{d\epsilon}{2\omega_d(2\omega_d+i2\gamma)-(2\epsilon)^2}\notag.
\end{align}
Performing the integral over the Lorentzian, we get:
\begin{equation}
    \chi_\Lambda^0(2\omega_d)  \simeq \int_{\omega_{\rm JP} }^{\infty} \, f_{\rm BG}(\epsilon,2\omega_d) \, d\epsilon  
     -\frac{i\pi \Lambda_{\rm eq}\mathcal{N}(\omega_d)\bar{\cal D}^{\psi\psi}_{\omega_d}}{8\sqrt{\omega_d(\omega_d+i\gamma)}}.
\end{equation}
We have thus shown that the response function $\chi^0_\Lambda$ can be decomposed into a well-defined background  term $\int^\infty_{\omega_{\rm JP}}f_{\rm BG}d\epsilon$, that only weakly depends on  $\omega_d$, and a term that scales with $\mathcal{N}(\omega_d)\times n_{\omega_d}$. A simple background subtraction should therefore isolate the finite momentum thermal fluctuations. 

\subsection{Third Harmonic Generation}
\begin{figure}
    \centering
    \includegraphics[width=\linewidth]{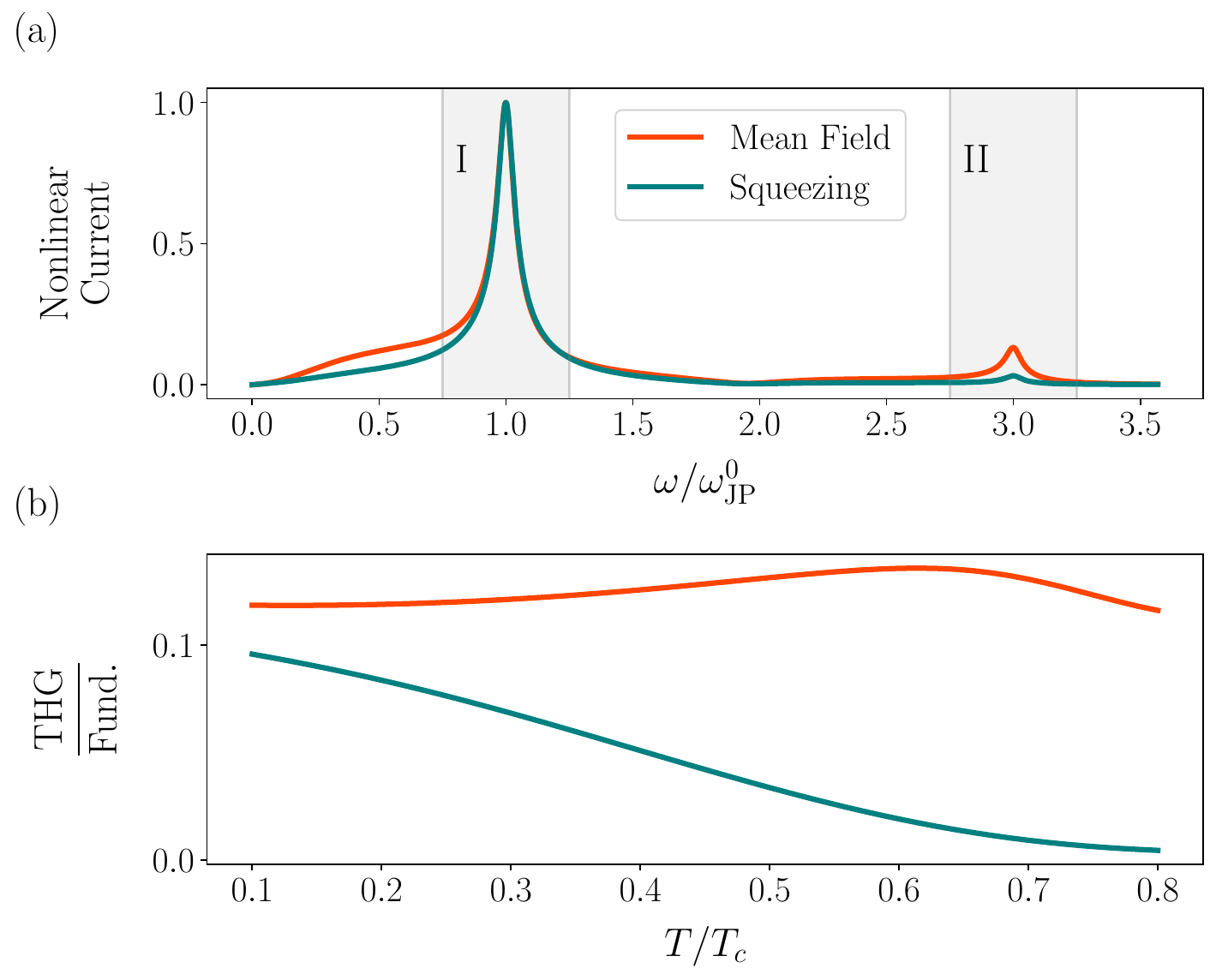}
    \caption{Third harmonic generation (THG). (a) Typical wave-mixing signals (here $T/T_c=0.5$) due to the mean-field and squeezing nonlinearities appear to have a substantial spectral overlap, which makes it difficult to isolate the squeezing channel. (b) Temperature evolution of the ratio of the third harmonic spectral intensity (integral under the shaded area II) to that of the fundamental one (integral under the shaded area I) shows that THG becomes less sensitive to thermally excited plasmons as $T$ approaches $T_c$. The driving frequency is fixed to $\omega_d/\omega_{\rm JP}^0 = 1$ for all temperatures. }
    \label{fig:7}
\end{figure}

One may also draw a comparison between 2DTS and third harmonic generation (THG), which probes a subset of the full nonlinearity accessible with 2DTS. THG is a powerful one-dimensional spectroscopic technique that has been extensively applied to study nonlinear optical properties of cuprates. 
Examples include investigation of striped superconductors with intertwined orders~\cite{rajasekaran2018probing}, origins of light-induced superconductivity~\cite{kim2023tracing}, out-of-plane Josephson plasmons~\cite{Katsumi_2023,Kaj_2023}, and Higgs mode signatures~\cite{cea2016nonlinear,chu2020phase}. 
In such experiments, photoexcitation consists of a narrowband pulse centered at the carrier frequency $\omega_d$, while the experimental observable is light emission at $3\omega_d$:
\begin{equation}
    E_{\rm signal}(t) \propto \chi^{(3)}(\omega_d,\omega_d,\omega_d)e^{-i3\omega_d t} + c.c.
\end{equation}
While THG gives partial access to $\chi^{(3)}$, we find that the contribution of squeezing nonlinearities to THG is difficult to separate from the mean-field one (see Fig.~\ref{fig:7}). Indeed, both nonlinearities have similar THG spectral profiles, while the sensitivity to thermally excited plasmons, as measured by the third harmonic spectral weight relative to the fundamental harmonic, decreases with temperature. This result is in full agreement with Ref.~\cite{Gabriele2021} and clearly illustrates the advantage of 2DTS in studying dynamical effects beyond mean-field both in the context of Josephson plasmonics and beyond.
\section{Conclusion and Outlook}

In this paper we developed a theory for 2DTS of the Josephson plasma resonance in layered superconductors. The nonlinear electrodynamics were cast in the form of optical susceptibility, from which explicit expressions for the 2D-spectra were obtained. 

For low temperatures, at which mean-field approximations hold, the spectra obtained remarkably follow intuition derived from the conventional scenario of quantum level systems \cite{Siemens_2010}. We demonstrated that the rephasing nonlinearity unambiguously separates homogeneous and inhomogeneous broadening in the impulsive limit, and that increasing temperature results in spectral peaks that directly follow the softening of the loss function. 

Near the phase transition however, the breakdown of the mean-field picture results in qualitative changes to the 2D-spectra that are both absent in the linear optical response and unique to collective excitations and their interactions. We found that non-mean-field corrections to the nonlinear susceptibility, such as the ones due to the parametric generation of squeezed plasmon pairs at finite momenta, manifest as distinct signatures in 2D-spectra and can be unambiguously isolated by proper choice of excitation pulse spectra. 
Our theory provides a natural interpretation to recent experimental observations~\cite{albert_phase_transition}, and paves the way for 2DTS to be an experimental probe of finite-momentum thermal fluctuations. By comparison, we found that third-harmonic generation \cite{Kaj_2023,kim2023tracing}, a one-dimensional nonlinear spectroscopy technique, is both ambiguous to the origin of the probed optical nonlinearity and also exhibits a decreasing sensitivity to plasma fluctuations. 

This comprehensive development of a theoretical framework for 2DTS of a collective excitation immediately motivates a new range of directions for both theoretical and experimental investigations. For example, many of the most enigmatic quantum materials feature coupling between several degrees of freedom \cite{Fradkin2015,Fernandes2019}, and theoretical exploration of how these couplings between various collective excitations manifest in 2DTS remains relatively unexplored. In this context, we are particularly interested in striped superconducting phases due to intertwined superconducting and charge (and possibly other) orders, which exhibit frustrated linear optical responses but have previously been shown to retain optical nonlinearities~\cite{rajasekaran2018probing}.

Such theoretical development could be particularly important in the context of understanding the phenomenon of light-induced superconductivity, interesting both fundamentally and technologically. In cuprates, coupling between Josephson plasmons and particular phonon modes~\cite{Liu2020,von2022amplification,Michael2022} has proven crucial to putative non-equilibrium superconductivity, but is often poorly understood. In other instances, metastable photo-induced superconductivity could originate from an interplay between competing orders such as superconductivity and charge order~\cite{Nicoletti2014,cremin2019photoenhanced}. Understanding how various order parameters and their associated collective modes couple will shed light on these intriguing phenomena, which calls for a 2DTS probe to disentangle the complex underlying physics.

Another important direction could be towards extending the presented model to incorporate optical cavities. Resonant cavities have been proposed to strongly modify the coupling of Josephson plasmons with light \cite{Laplace2016,Kalhor2022} in both interesting and functional ways, and these changes should manifest in 2D-spectra as well \cite{Xiang2018,Xiang2020}. Beyond the Josephson plasma resonance, this is also motivated by recent experiments demonstrating spectacular control of a phase transition using a terahertz cavity~\cite{Jarc2023}. The specific mechanisms leading to these effects remain uncertain however, which also calls for a 2DTS probe to clarify the underlying physics.

More broadly, we believe this work to be a valuable first step towards a general theoretical description for 2DTS of complex solids, and expect many of the results derived here for the Josephson plasma resonance to apply generically for 2DTS of collective excitations. Phonons~\cite{Teitelbaum2018}, magnons~\cite{Jain2017}, and charge density waves~\cite{Liu2013} are but a few examples of such collective excitations for which nonlinearities play an essential role, and 2DTS should provide invaluable insight into these important systems that were previously accessible only via high-energy probes. The general principles we demonstrate here for a model collective excitation will further guide the design of future 2DTS experiments in targeting the most striking effects of quantum materials.

\section*{ACKNOWLEDGEMENTS}
The authors would like to thank J. Curtis, L. Glazman, S. Gopalakrishnan, E. Kaxiras, P.A. Lee, F. Marijanovic, M. Mitrano, and D. Nicoletti  for fruitful discussions. A.G.S and E.D. acknowledge the Swiss National Science Foundation (Grants Number 200021 and 212899) and ETH-C-06 21-2 equilibrium Grant with project number 1-008831-001 for funding. The work of P.E.D. was sponsored by the Army Research Office and was accomplished under Grant Number W911NF-21-1-0184. A.L. was supported by the U.S. Department of Energy, Office of Basic Energy Sciences, under Contract No. DE- SC0012704.


%


\newpage
\clearpage
\onecolumngrid
\appendix
\section{Generalized reflectivity protocol}
\label{appendix:nonlinear_Gen}
This Appendix discusses that our derivations  of Josephson plasmon nonlinearities in Sec.~\ref{section_2} can be straightforwardly extended to other types of collective modes. We begin by writing down Maxwell's equations in the sample~\cite{Mukamel}
\begin{gather}
    \nabla D = 0, \qquad \nabla B = 0, \qquad
    \nabla \times E = -\frac{1}{c}\frac{\partial B}{\partial t}, \qquad \nabla\times H=\frac{1}{c}\frac{\partial D}{\partial t},
\end{gather}
alongside with the constitutive relations
\begin{gather}
    D = E +4\pi P, \qquad H = B -4\pi M , \qquad J = \frac{\partial P}{\partial t} + c \nabla \times M.
\end{gather}
Taking the curl of Faraday's law, we arrive at
\begin{gather}
    \nabla\times\nabla\times E +\frac{1}{c^2}\frac{\partial^2E}{\partial t^2} = -\frac{4\pi}{c^2}\frac{\partial J}{\partial t}.
    \label{El_Pol}
\end{gather}
If we are interested only in the polarization response of the system ($M=0$), for the transverse component of the electric field we then get
\begin{equation}
    \nabla^2 E^{\perp} -\frac{1}{c^2}\frac{\partial^2E^{\perp}}{\partial t^2} = \frac{4\pi }{c^2}\frac{\partial^2 P^{\perp}}{\partial t^2}.
\end{equation}
Here, the polarization vector acts as a source for the electric field. We now split the polarization vector into its linear part, proportional to the electric field, and non-linear part, containing higher orders of the electric field: $P = P^{(1)} + P_{\text{NL}} $. The linear relation between $P^{(1)}$ and $E(r',t')$ can be expressed via the response function  $\chi(r-r',t-t')$ as
\begin{equation}
    P^{(1)}(r,t)=\int d^d r' \int_0^t dt' \chi(r-r',t-t')E(r',t').
\end{equation}
Plugging into (\ref{El_Pol}) and defining the dielectric function as $\varepsilon(r-r',t-t') = \delta(t-t')\delta(r-r')+4\pi\chi(t-t',r-r')$, we arrive at
\begin{equation}
    \nabla\times\nabla\times E+\frac{1}{c^2}\frac{\partial^2}{\partial t^2}\int d^dr'\int_0^t dt'\, \varepsilon(r-r',t-t')E(r',t')=-\frac{4\pi}{c^2}\frac{\partial^2 P_{\text{NL}}}{\partial t^2}
\end{equation}
and
\begin{gather}
    \nabla^2 E^{\perp} -\frac{1}{c^2}\frac{\partial^2}{\partial t^2} \int d^dr'\int_0^t dt'\, \left[\varepsilon(r-r',t-t')E(r',t')\right]^\perp = \frac{4\pi }{c^2}\frac{\partial^2 P^{\perp}_{\text{NL}}}{\partial t^2}.
\end{gather}

The nonlinear polarization $P_{\text{NL}}$ carries the complete microscopic information describing any nonlinear optical process of interest; $P_{\text{NL}}$ is to be obtained from the microscopic dynamics of the system. Restricting ourselves to the transverse sector and henceforth dropping $\perp$, we can rewrite the previous equation in terms of the vector potential as
\begin{gather}
    \nabla^2 A -\frac{1}{c^2}\frac{\partial^2}{\partial t^2} \int d^dr'\int_0^t dt'\, \varepsilon(r-r',t-t')A(r',t')= -\frac{4\pi }{c}\frac{\partial P_{\text{NL}}}{\partial t}.
    \label{eqn: A_material_reflect}
\end{gather}
Assuming normal incidence, the Fresnel boundary conditions are written as:
\begin{align}
      E^{\rm in}_z(y,t) + E^{\rm r}_z(y,t)  = E^{\rm t}_z(y,t) = -\frac{1}{c}\partial_t A(\boldsymbol{r},t)\big{|}_{x=0}, \qquad     
    B_y^{\rm in}(y,t) + B_y^{\rm r}(y,t)  = B_y^{\rm t}(y,t)=-\partial_x A(\boldsymbol{r},t)\big{|}_{x=0}.
\end{align}
Equation~\eqref{eqn:EB_relations_air} further yields:
\begin{equation}
    2 E^{\rm in}(y,t)=\Big(\partial_x A(\boldsymbol{r},t) - \frac{1}{c}\partial_t A(\boldsymbol{r},t)\Big)\Bigg{|}_{x=0} \text{ and } E^{\rm r}(y,t) = -\frac{1}{2}\Big(\partial_x A(\boldsymbol{r},t) + \frac{1}{c}\partial_t A(\boldsymbol{r},t)\Big)\Bigg{|}_{x=0}.
\end{equation}
The presented equations are generic and represent the starting point for analysing nonlinearities of many-body systems.

Following the derivation steps in Sec.~\ref{section_2}, we now illustrate how such nonlinearities could be evaluated in practice using as an example  the nonlinear Lorentz oscillator model:
\begin{align}
    \left(\partial_t^2  +\Gamma \partial_t - v^2\partial_x^2 + m^2\right) P(x,t) + g  P^3(x,t) = E(x,t).
\end{align}
(Note that here, in order for $P$ and $E$ to have the same unit, frequencies have been rescaled to be dimensionless.) We now expand the polarization vector in powers of the electric field $P(x,t) = P^{(1)}(x,t) +P^{(3)}(x,t)+\dots$: 
\begin{align}
    \left(\partial_t^2 +\Gamma \partial_t - v^2\partial_x^2 + m^2 \right)P^{(1)}(x,t) &= E(x,t), \label{eqn: P_lin}\\
    \left(\partial_t^2 +\Gamma \partial_t - v^2\partial_x^2 + m^2 \right)P^{(3)}(x,t) &= -g \Big(P^{(1)}(x,t)\Big)^3.
    \label{eqn: P_nl}
\end{align}
Eq.~\eqref{eqn: P_lin} yields the response function and dielectric function to be given by:
\begin{gather}
    \chi(k,\omega) = \frac{-1}{\omega(\omega+i\Gamma)-v^2k^2-m^2},  \qquad
    \epsilon( k, \omega) = 1- \frac{4\pi}{\omega(\omega+i\Gamma)-v^2k^2-m^2}. \label{eqn: diel_func_P}
\end{gather}
The Fresnel boundary conditions for the vector potential then result in:
\begin{align}
    \Big(\partial_x A^{(1)}(x,t) - \frac{1}{c}\partial_t A^{(1)}(x,t)\Big)\Bigg{|}_{x=0} = 2 E^{\rm in}(0,t),\qquad
    \Big(\partial_x A^{(3)}(x,t) - \frac{1}{c}\partial_t A^{(3)}(x,t)\Big)\Bigg{|}_{x=0} = 0. \label{eqn: BC A3}
\end{align}

The linear response can be related to the dielectric function as:
\begin{equation}
    A^{(1)}(x,t)=\int\frac{d\omega}{2\pi} \tilde{A_1}(\omega)e^{ik_x(\omega)x-i\omega t},
\end{equation}
where $k_x(\omega)$ is defined through 
\begin{equation}
    k_x^2 = \frac{\omega^2}{c^2}\epsilon(k_x,\omega)\approx\frac{\omega^2}{c^2}\epsilon(k=0,\omega)\equiv \frac{\omega^2}{c^2}\epsilon(\omega) .
\end{equation}
This approximation is justified whenever $v\ll c$, which is expected to hold for a typical solid-state system. The amplitudes $\tilde{A}^{(1)}(\omega)$ are determined from the boundary conditions~\eqref{eqn: BC A3}:
\begin{equation}
    \tilde{A}^{(1)}(\omega) = \frac{c}{i\omega}t(\omega)\tilde{E}^{\rm in}(\omega), \text{ where } t(\omega)=\frac{2}{1+\sqrt{\epsilon(\omega)}}.
\end{equation}
The linear polarization is then given by:
\begin{equation}
    P^{(1)}(x,t) \simeq \int \frac{d\omega}{2\pi}\chi(\omega)t(\omega)E^{\rm in}(\omega)e^{ik_x(\omega)x-i\omega t},
\end{equation}
where again we have assumed $v\ll c$ and defined $\chi(\omega)\equiv\chi(k = 0,\omega)$. 

Having determined the leading harmonic $P^{(1)}(x,t)$, we turn to compute $P^{(3)}(x,t)$. Since Eq.~\eqref{eqn: P_nl} is a linear differential equation, we easily obtain:
\begin{align}
    P^{(3)}(x,t) & = -g\int \frac{d\omega}{2\pi}\int \frac{d k}{2\pi} \, \chi(\omega) e^{ikx - i\omega t} \int \frac{d\omega_1}{2\pi}\frac{d\omega_2}{2\pi}\frac{d\omega_3}{2\pi} 2\pi\delta(\omega_1 + \omega_2 + \omega_3 - \omega)  2\pi\delta(k_x(\omega_1) + k_x(\omega_2) + k_x(\omega_3) - k) \notag\\
    &
    \times \left[\chi(\omega_1)t(\omega_1)E^{\rm in}(\omega_1)\right]\left[\chi(\omega_2)t(\omega_2)E^{\rm in}(\omega_2)\right]\left[\chi(\omega_3)t(\omega_3)E^{\rm in}(\omega_3)\right].
\end{align}
The equation on $A^{(3)}(x,t)$ is also a linear differential equation, and therefore we can write the generic solution as:
\begin{equation}
    A^{(3)}(x,t) = \int \frac{d\omega}{2\pi}\Tilde{A}^{(3)}(\omega) e^{ik_x(\omega)x -i\omega t} + A^{(3)}_{\rm dr}(x,t),
\end{equation}
where 
\begin{equation}
    A^{(3)}_{\rm dr} (x,t) = 4\pi ic \int \frac{dk}{2\pi} \int \frac{d\omega}{2\pi}\frac{\omega P^{(3)}(k,\omega)}{\omega^2\epsilon(k,\omega)-k^2c^2}e^{-i(\omega t - k x)}.
\end{equation}
The coefficients $\tilde{A}^{(3)}(\omega)$ are determined from the boundary condition~\eqref{eqn: BC A3}:
\begin{equation}
    \tilde{A}^{(3)}(\omega)=-\int\frac{dk}{2\pi}\frac{\omega+ck}{\omega+ck_x(\omega)}A^{(3)}_{\rm dr}(k,\omega).
\end{equation}
For the ease of notations, we define $\bar{\omega} = \omega_1+\omega_2+\omega_3$ and $k_x(\omega_1,\omega_2,\omega_3) = k_x(\omega_1) + k_x(\omega_2) + k_x(\omega_3)$.
For the reflected light at $x =0$, we finally obtain:
\begin{align}
    E^{\rm r}_3(t)&  = -\frac{1}{2}\Big(\frac{1}{c}\partial_t A^{(3)}(0,t) + \partial_x A^{(3)}(0,t)\Big) = -i\int\frac{d\omega}{2\pi}\int\frac{dk}{2\pi}\frac{\omega\left[k-k_x(\omega)\right]}{\omega+k_x(\omega)} A^{(3)}_{\rm dr}(k,\omega)e^{-i\omega t}   \notag \\ & 
    = \int\frac{d\omega_1}{2\pi}\int\frac{d\omega_2}{2\pi}\int\frac{d\omega_3}{2\pi}  e^{-i\bar{\omega} t} \chi^{(3)}(\omega_1,\omega_2,\omega_3) E^{\rm in}(\omega_1)E^{\rm in}(\omega_2)E^{\rm in}(\omega_3).
\end{align}
so that the third-order susceptibility is then given by:
\begin{align}
    \chi^{(3)}(\omega_1,\omega_2,\omega_3) &= \frac{2\pi g}{c k_x(\omega_1,\omega_2,\omega_3)/\bar{\omega} + \sqrt{\epsilon(\bar{\omega})} } \left[\chi(\bar{\omega})t(\bar{\omega})\right]\left[\chi(\omega_1)t(\omega_1)\right] \left[\chi(\omega_2)t(\omega_2)\right] \left[\chi(\omega_3)t(\omega_3)\right].   \label{eqn:chi_3_gen_appendix}
\end{align}
\section{Equilibrium fluctuations within the Gaussian approximation} \label{app: equilbirum}

In this Appendix, we discuss the equilibrium correlation functions obtained by self-consistently solving Eqs.~\eqref{eqn: eql} and~\eqref{eqn: Lambda_eq}. To this end, it is natural to consider the following quantity:
\begin{equation}
    \Omega(\tilde{T}, {\cal C})\equiv\frac{1}{{\cal A}N}\sum_{\bm q, q_z} 
    \bar{\mathcal{D}}_{\bm q, q_z}^{\psi \psi} = \int_{-\pi}^\pi \frac{dq_z}{2\pi}\int_0^{\cal C}\frac{dq}{2\pi}\frac{q\,\tilde{T}}{q^2 L(q_z) + \Lambda_{\rm eq}},
\end{equation}
where ${\cal C}$ is the UV momentum cutoff. The in-plane integration results in
\begin{align}
    \Omega(\tilde{T};\mathcal{C}) & =\frac{\tilde{T}}{8\pi^2}\int_{-\pi}^\pi dq_z \frac{1}{L(q_z)}\log\Big(1+\frac{\mathcal{C}^2L(q_z)}{\Lambda_{\rm eq}}\Big)  \simeq \frac{\tilde{T}}{8\pi^2}\int_{-\pi}^\pi dq_z \frac{1}{L(q_z)}\log
    \Big(\frac{\mathcal{C}^2L(q_z)}{\Lambda_{\rm eq}}\Big),
\end{align}
where we have assumed that $\mathcal{C}\gg\sqrt{\Lambda_{\rm eq}}/\min_{q_z}(L(q_z))$. 
The integral over $q_z$ can also be evaluated explicitly:
\begin{align}
    \Omega(\tilde{T};\mathcal{C}) = \frac{\tilde{T}}{4\pi c_0^2}\Big{\{} &b\Big[ \log \Big(\frac{2c_0^2\mathcal{C}^2/\Lambda_{\rm eq}}{b+\sqrt{b^2-4}} \Big)-1\Big] + \sqrt{b^2-4} \Big{\}},
\end{align}
where we have defined $b\equiv 2+s^2/\lambda_{ab}^2$. Substituting $\Lambda_{\rm eq} = \Lambda_0 \exp(-\Omega/2)$, we arrive at:
\begin{align}
    \Omega(\tilde{T},\mathcal{C}) = \frac{\tilde{T}/4\pi c_0^2}{1-\Tilde{T}b/8\pi c_0^2}\Big{\{} & b\Big[\log \Big(\frac{2c_0^2\mathcal{C}^2/\Lambda_{0}}{b+\sqrt{b^2-4}}\Big)-1\Big] +\sqrt{b^2-4} \Big{\}}. \label{eqn: Omega_self}
\end{align}
From this result, we obtain the critical temperature, i.e., where the Josephson plasmon resonance softens to zero:
\begin{equation}
    \Tilde{T}_c = \frac{8\pi c_0^2}{2+ s^2/\lambda_{ab}^2} = \frac{8\pi c^2 s^2/\lambda_{ab}^2}{\epsilon_\infty\left(2+ s^2/\lambda_{ab}^2\right)}.
\end{equation}
The central result of this Appendix is Eq.~\eqref{eqn: Omega_self}, which can be used to obtain any other equilibrium observable of interest.

\section{Dynamics of two-point correlators} \label{Appendix: dynamics of correlators}

Considering the perturbation given in Eq.~\eqref{eqn: delta_Lambda}, the linearized equations of motion for the two-point correlators around the equilibrium state are given by:
\begin{gather}
    -i\omega \delta{\cal D}^{\psi\psi}_{\bm q, q_z}   =  2\delta{\cal D}^{\psi\pi}_{\bm q, q_z} , \label{eqn:D1}\\
    -i(\omega +i \gamma) \delta{\cal D}^{\psi\pi}_{\bm q, q_z}   = \delta{\cal D}^{\pi\pi}_{\bm q, q_z} - [\bm q^2 L(q_z) + \Lambda_{\rm eq}]\delta{\cal D}^{\psi\psi}_{\bm q, q_z} 
    -\Lambda_{\rm eq} \Big[ 
     \frac{\delta \Lambda_0}{\Lambda_0}  - \frac{1}{2 {\cal A} N} \sum_{\bm k,k_z}
     \delta{\cal D}_{\bm k,k_z}^{\psi\psi}
    \Big]\bar{\mathcal{D}}_{\bm q,q_z}^{\psi\psi} ,\label{eqn:D2}\\
    -i(\omega+2i\gamma) \delta{\cal D}^{\pi\pi}_{\bm q, q_z} = - 2 [\bm q^2 L(q_z) + \Lambda_{\rm eq} ]\delta{\cal D}^{\psi\pi}_{\bm q, q_z}.\label{eqn:D3}
\end{gather}
We observe that the last term in Eq.~\eqref{eqn:D2} mixes different momenta modes, suggesting to split the propagator of fluctuations into the diagonal part
\begin{gather}
        G_0^{-1}(\omega;q,q') =  \begin{pmatrix} -i\omega & -2 & 0 \\ 
        \bm q ^2L(q_z) +\Lambda_{\rm eq}  & -i(\omega + i\gamma) & -1\\ 0 & 2[\bm q^2L(q_z) + \Lambda_{\rm eq}] & -i(\omega + 2i\gamma) \end{pmatrix}\delta_{q,q'}
\end{gather}
and the off-diagonal part
\begin{align}
    \mathcal{V}(q,q') = -\frac{\Lambda_{\rm eq}\bar{\mathcal{D}}_{q}^{\psi\psi}}{2{\cal A}N}\begin{pmatrix} 0 & 0 & 0 \\
     1&0 & 0\\ 0 & 0&0 \end{pmatrix},
\end{align}
where
\begin{align}
    \bm {\delta\mathcal{D}}_{q} = \begin{pmatrix} \delta\mathcal{D}_{q}^{\psi\psi},\delta\mathcal{D}_{q}^{\psi\pi},\delta\mathcal{D}_{q}^{\pi\pi} \end{pmatrix}^\top,
    \qquad
    \label{eqn: d-correl basis}
     q = \begin{pmatrix} \bm q, q_z \end{pmatrix}.
\end{align}
We also define a driving vector $\bm f$ as:
\begin{equation}
    \boldsymbol {f}_q(\omega) = -\bar{\mathcal{D}}_{q}^{\psi\psi}\Lambda_{\rm eq} \frac{\delta\Lambda_0(\omega) }{\Lambda_{0}}\begin{pmatrix} 0 \\ 1 \\ 0 \end{pmatrix}.
\end{equation}

One can formally write the solution of Eqs.~\eqref{eqn:D1}-\eqref{eqn:D3} as follows:
\begin{align}
    \bm {\delta\mathcal{D}}_k &= \sum_{k'}\left( \frac{1}{G_0^{-1}+\mathcal{V}} \right)_{k,k'}\boldsymbol f_{k'} = (G_0\boldsymbol{f})_k - \left( G_0\mathcal{V}G_0\boldsymbol{f} \right)_k + \left( G_0\mathcal{V}G_0\mathcal{V}G_0\boldsymbol{f} \right)_k- \dots \notag \\
    &= \sum_{k'}G_0(k,k')\boldsymbol{f}_{k'} - \sum_{p}G_0(k,p)\sum_{k''}\mathcal{V}(p,k'')\sum_{k'}G_0(k'',k')\boldsymbol{f}_{k'}+\dots 
\end{align}
Since $\mathcal{V}(k,q)$ depends only on $k$ and $G_0(k,k')\propto \delta_{k,k'}$, we further have:
\begin{align}
     \bm {\delta\mathcal{D}}_k &= G_0(k)\boldsymbol{f}_{k} - G_0(k)\mathcal{V}(k)\sum_{k'}G_0(k')\boldsymbol{f}_{k'} + G_0(k)\mathcal{V}(k)\sum_{k''}G_0(k'')\mathcal{V}(k'')\sum_{k'}G_0(k')\boldsymbol{f}_{k'}-\dots
\end{align}
Summing both sides with respect to $k$, we get:
\begin{align}
    \frac{1}{{\cal A}N}\sum_{\bm k,k_z}\bm {\delta\mathcal{D}}_k &= \Big( 1 - \sum_{p}G_0(p)\mathcal{V}(p) + \Big(\sum_{p}G_0(p)\mathcal{V}(p)\Big)^2 +\dots \Big)\frac{1}{{\cal A}N}\sum_{k'}G_0(k')\boldsymbol{f}_{k'}  \notag \\
    &= 
    \Big( \frac{1}{1+\sum_{p}G_0(p)\mathcal{V}(p)}\Big) \frac{1}{{\cal A}N}\sum_{k}G_0(k)\boldsymbol{f}_{k}.\label{eqn: resumm}
\end{align}
In particular, for the $\psi\psi$-component, Eq.~\eqref{eqn: resumm} yields:
\begin{gather}
    \frac{1}{2{\cal A}N}\sum_{\bm q, q_z} \delta\mathcal{D}^{\psi\psi}_{\bm q,q_z}(\omega) = - \frac{\chi^0_\Lambda(\omega)}{1- \chi^0_\Lambda(\omega)}  \frac{\delta\Lambda_0(\omega) }{\Lambda_{0}} = -\chi_\Lambda(\omega) \frac{\delta\Lambda_0(\omega) }{\Lambda_{0}},
    \label{eqn: def_chi_lambda}
\end{gather}
where
\begin{align}
    \chi^0_\Lambda(\omega)  & = - \frac{\omega+2i\gamma}{\omega+i\gamma}\frac{1}{{\cal A}N}\sum_{\bm q, q_z} \frac{\bar{\mathcal{D}}_{\boldsymbol q,q_z}^{\psi\psi}\Lambda_{\rm eq}}{\omega (\omega + 2i\gamma) - 4[\bm q^2 L(q_z) + \Lambda_{\rm eq} ]}\label{eqn: chi_int} \\
    & = - \frac{2\tilde{T}\Lambda_{\rm eq}}{\tilde{T}_c\omega(\omega+i\gamma)} \log\left( 1-i\frac{\gamma\omega}{2\Lambda_{\rm eq}}-\frac{\omega^2}{4\Lambda_{\rm eq}} \right)
    \label{eqn: integral}.
\end{align}
This completes our derivation of Eqs.~\eqref{eqn:chi_Lambda_v0} and~\eqref{eqn:chi_Lambda_v1} of the main text.
\section{The third order response function and plasma squeezing}
\label{appendix:3rd order}

Following Sec.~\ref{section_2}, here we evaluate the third order response function but now take into account the dynamics of the  electromagnetic background. To mimic an incoming train of light pulses, we consider current perturbations $j_z(t)\sim E^{\rm in}_z(t)$, Eq.~\eqref{eqn:dyn_1p}, as well as perturabtions of the bare coupling constant $\delta \Lambda_0(t) \sim [E^{\rm in}_z(t)]^2$, Eq.~\eqref{eqn: delta_Lambda}. Explicit expansion to the third order in the incoming electric field acquires the form:
\begin{gather}
    \psi(t) = \psi^{(1)}(t) + \psi^{(3)}(t) + \cdots , \\
    \pi(t)  = \pi^{(1)}(t) + \pi^{(3)}(t) + \cdots , \\
    \mathcal{D}_{\bm q,q_z}^{\alpha\beta}(t) = \bar{\mathcal{D}}_{\bm q,q_z}^{\alpha\beta}(t) + \mathcal{D}_{\bm q,q_z}^{\alpha\beta(2)}(t) + \cdots\,.
\end{gather}
Direct insertion of this expansion into Eqs.~\eqref{eqn:dyn_1p}-\eqref{eqn:dyn_2p_v3} gives:
\begin{equation}
    \frac{\psi^{(3)}(\omega)}{\chi_\psi(\omega)} =\frac{\Lambda_{\rm eq}}{6}\mathcal{FT}\Big[ \Big(\psi^{(1)}(t)\Big)^3 \Big] -\mathcal{FT}\Big[ \Big(\Lambda_{\rm eq}\frac{\delta\Lambda_0(t)}{\Lambda_0}-\frac{\Lambda_{\rm eq}}{2\mathcal{A}N}\sum_{\bm q,q_z}
    {\cal D}_{\bm q,q_z}^{\psi\psi(2)}(t)\Big)\psi^{(1)}(t) \Big], 
    \label{eqn:psi_3_sq_v0}
\end{equation}
where $\psi^{(1)}(\omega) = \chi_\psi(\omega)j_z(\omega)$ and $\mathcal{FT}$ stands for the Fourier transform. The dynamics of $\mathcal{D}^{\psi\psi(2)}_{\bm q,q_z}$ follows from linearizing Eqs.~\eqref{eqn:dyn_2p_v1}-\eqref{eqn:dyn_2p_v3} on top of the equilibrium state:
\begin{gather}
    -i\omega{\cal D}^{\psi\psi(2)}_{\bm q, q_z}   =  2{\cal D}^{\psi\pi (2)}_{\bm q, q_z} , 
    \label{eqn:D1_v2}\\
    -i(\omega +i \gamma) {\cal D}^{\psi\pi(2)}_{\bm q, q_z}   = {\cal D}^{\pi\pi(2)}_{\bm q, q_z} - [\bm q^2 L(q_z) + \Lambda_{\rm eq}]{\cal D}^{\psi\psi(2)}_{\bm q, q_z} 
    -\Lambda_{\rm eq} \Big[ 
     \frac{\delta \Lambda_0}{\Lambda_0}  - \frac{1}{2 {\cal A} N} \sum_{\bm k,k_z}
     {\cal D}_{\bm k,k_z}^{\psi\psi(2)}
     -\frac{1}{2} (\psi^{(1)}*\psi^{(1)}) (\omega)
    \Big]\bar{\mathcal{D}}_{\bm q,q_z}^{\psi\psi} ,\label{eqn:D2_v2}\\
    -i(\omega+2i\gamma) {\cal D}^{\pi\pi(2)}_{\bm q, q_z} = - 2 [\bm q^2 L(q_z) + \Lambda_{\rm eq} ]{\cal D}^{\psi\pi (2)}_{\bm q, q_z},\label{eqn:D3_v2}
\end{gather}
where $*$ denotes convolution. The only difference between Eqs.~\eqref{eqn:D1_v2}-\eqref{eqn:D3_v2} and Eqs.~\eqref{eqn:D1}-\eqref{eqn:D3} is that the left hand side of Eq.~\eqref{eqn:D2_v2} contains an additional driving term $\propto (\psi^{(1)}(t))^2$. Employing Eq.~\eqref{eqn: def_chi_lambda}, we, thus, readily obtain:
\begin{equation}
    - \frac{\Lambda_{\rm eq}}{2\mathcal{A}N}\sum_{\bm q,q_z}\mathcal{D}^{\psi\psi(2)}_{\bm q,q_z}(\omega) =\chi_\Lambda(\omega) \Lambda_{\rm eq}\left[ \frac{\delta\Lambda_0(\omega) }{\Lambda_{0}}-\frac{ (\psi^{(1)}*\psi^{(1)}) (\omega)}{2}\right].
\end{equation}

Equation~\eqref{eqn:psi_3_sq_v0} then yields:
\begin{equation}
    \frac{\psi^{(3)}(\omega)}{\chi_\psi(\omega)} = \frac{\Lambda_{\rm eq}}{6}\mathcal{FT}\Big[ \Big(\psi^{(1)}(t)\Big)^3 \Big] 
    - \mathcal{FT}\Big\{ \mathcal{FT}^{-1}\Big[\Big(1+\chi_\Lambda(\omega)\Big)\Lambda_{\rm eq}\frac{\delta\Lambda_0(\omega)}{\Lambda_0}-\chi_\Lambda(\omega)\Lambda_{\rm eq}\frac{ (\psi^{(1)}*\psi^{(1)}) (\omega)}{2}\Big]\psi^{(1)}(t) \Big\}.
\end{equation}
The third order dynamics of $\psi$ can be written as:
\begin{align}
    \psi^{(3)}(t) &= \int\frac{d\omega_1}{2\pi}\int\frac{d\omega_2}{2\pi}\int\frac{d\omega_3}{2\pi}  e^{-i(\omega_1+\omega_2+\omega_3)t} \chi^{(3)}(\omega_1,\omega_2,\omega_3) j_z(\omega_1)j_z(\omega_2)j_z(\omega_3)\notag \\ &+ \int\frac{d\omega_1}{2\pi}\int\frac{d\omega_2}{2\pi}e^{-i(\omega_1+\omega_2)t}\tilde{\chi}^{(3)}(\omega_1,\omega_2)\delta\Lambda_0(\omega_1)j_z(\omega_2),
\end{align}
where $\chi^{(3)}(\omega_1,\omega_2,\omega_3)=\chi^{(3)}_{\rm mf}(\omega_1,\omega_2,\omega_3) + \chi^{(3)}_{\rm sq}(\omega_1,\omega_2,\omega_3)$ with:
\begin{align}
    \chi_{\rm mf}^{(3)}(\omega_1,\omega_2,\omega_3) &= \frac{\Lambda_{\rm eq}}{6}\chi_\psi(\omega_1+\omega_2+\omega_3)\chi_\psi(\omega_1)\chi_\psi(\omega_2)\chi_\psi(\omega_3),\notag \\
    \chi_{\rm sq}^{(3)}(\omega_1,\omega_2,\omega_3) &= \frac{\Lambda_{\rm eq}}{6}\sum_{i = 1,2,3} \chi_\psi(\omega_1+\omega_2+\omega_3)\chi_\Lambda(\omega_1+\omega_2 + \omega_3 - \omega_i)\chi_\psi(\omega_1)\chi_\psi(\omega_2)\chi_\psi(\omega_3),  
\end{align}
and 
\begin{equation}
    \Tilde{\chi}^{(3)}(\omega_1,\omega_2)= -\frac{\Lambda_{\rm eq}}{\Lambda_0}\chi_\psi(\omega_1+\omega_2)\left[1+\chi_\Lambda(\omega_1)\right]\chi_\psi(\omega_2).
\end{equation}
This completes our derivations of Eqs.~\eqref{eqn:chi_3_mf} and \eqref{eqn:chi_3_sq} of the main text. 

\end{document}